\documentclass[aps,reprint,groupedaddress,superscriptaddress]{revtex4-2}

\usepackage{lipsum}

\usepackage[svgnames]{xcolor}
\usepackage[colorlinks=true,
            linkcolor=red,
            urlcolor=blue,
            citecolor=DarkGreen]{hyperref}

\usepackage{graphicx}

\usepackage{dcolumn}
\usepackage{bm}
\usepackage{amsmath,amssymb,amstext,mathtools}
\usepackage{physics}
\usepackage{siunitx}
\usepackage{csquotes}

\usepackage[capitalize]{cleveref}
\usepackage{natbib}
\usepackage{dsfont}
\usepackage{embedfile}

\begin{document}

\title{Fast quantum control of cavities using an improved protocol without coherent errors}


\author{Jonas~Landgraf}\email{Jonas.Landgraf@mpl.mpg.de}
\affiliation{Max Planck Institute for the Science of Light, Staudtstr.~2, 91058 Erlangen, Germany}
\affiliation{Physics Department, University of Bayreuth, Universitätsstr.~30, 95447 Bayreuth, Germany}
\affiliation{Physics Department, University of Erlangen-Nuremberg, Staudstr.~5, 91058 Erlangen, Germany}
\author{Christa~Fl\"uhmann}
\affiliation{Department of Applied Physics and Physics, Yale University, New Haven, Connecticut 06511, USA}
\affiliation{Yale Quantum Institute, Yale University, New Haven, Connecticut 06511, USA}
\author{Thomas~Fösel}
\author{Florian~Marquardt}
\affiliation{Max Planck Institute for the Science of Light, Staudtstr.~2, 91058 Erlangen, Germany}
\affiliation{Physics Department, University of Erlangen-Nuremberg, Staudstr.~5, 91058 Erlangen, Germany}
\author{Robert~J.~Schoelkopf}
\affiliation{Department of Applied Physics and Physics, Yale University, New Haven, Connecticut 06511, USA}
\affiliation{Yale Quantum Institute, Yale University, New Haven, Connecticut 06511, USA}

\date{\today}

\begin{abstract}

The selective number-dependent arbitrary phase (SNAP) gates form a powerful class of quantum gates, imparting arbitrarily chosen phases to the Fock states of a cavity. However, for short pulses, coherent errors limit the performance. Here we demonstrate in theory and experiment that such errors can be completely suppressed, provided that the pulse times exceed a specific limit. The resulting shorter gate times also reduce incoherent errors. Our approach needs only a small number of frequency components, the resulting pulses can be interpreted easily, and it is compatible with fault-tolerant schemes.

\end{abstract}

\maketitle


The field of circuit quantum electrodynamics (cQED) \cite{blais2021circuit} employs microwave cavities coupled to superconducting qubits \cite{wendin-review,krantz2019quantum} and defines one of the most promising platforms for quantum computation. Modern 3D cavities offer long coherence times up to milliseconds \cite{reagor2016quantum} and beyond \cite{romanenko2020three, milul2023superconducting}. This provides the opportunity to store and process the quantum information in the bosonic modes of the cavity \cite{gottesman2001encoding,mirrahimi2014dynamically,michael2016new,joshi2021quantum}. The strong coupling to the superconducting qubit allows fast and flexible manipulation of the cavity's quantum state \cite{blais2004cavity,wallraff2004strong,ma2021quantum}. These setups showed remarkable success in quantum simulations \cite{hu2018simulation,wang2020efficient,cai2021high}, quantum teleportation \cite{chou2018deterministic}, quantum state transfer \cite{axline2018demand} or implementing bosonic quantum error correction \cite{ofek2016,hu2019quantum,campagne2020quantum,gertler2021protecting,grimsmo2021quantum,sivak2022real}, even reaching the so-called break-even point.

Two different approaches exist to gain universal control over the cavity-qubit system. For the first one, cavity and qubit are driven simultaneously with pulse sequences that are typically numerically optimized \cite{GRAPE-cQED}, e.g. with GRAPE \cite{GRAPE-NMR}, to approximate a certain unitary operation. The second approach uses the powerful selective number-dependent arbitrary phase (SNAP) gate, which can impart any desired set of phases on the Fock states of the cavity \cite{heeres2015cavity}. For example, with cavity displacements, the SNAP gate can be easily extended to a universal gate set, for which \cite{krastanov2015universal,fosel2020efficient} provide efficient schemes to approximate any desired unitary. Because any errors accumulate in such sequences, it is crucial to improve the fidelity of an individual SNAP gate as far as possible to enable the realization of more complex unitaries. 

To both avoid incoherent errors and ensure a rapid overall processing speed, it is desirable to make the SNAP gate time as short as possible. However, in that regime, the fidelity suffers from coherent errors. 

To overcome this challenge, recently numerical techniques were employed to optimize the envelope of the SNAP pulses, providing fast SNAP gates and enable the preparation of arbitrary cavity states with high fidelity \cite{kudra2022robust}. However, numerical optimization schemes, like GRAPE \cite{GRAPE-cQED,GRAPE-NMR}, tend to have thousands of adjustable parameters, are thus hard to interpret, and incorporate high-frequency components. In our work, we present a simple approach which, above some pulse duration, can completely suppress the coherent errors based on a geometrical interpretation of the errors. Our optimized pulses are continuous in time and preserve the original form of the SNAP gate pulses \cite{heeres2015cavity} without introducing additional frequencies. We demonstrate experimentally that our pulses reduce the excited state error of the SNAP gate, the dominant coherent error, by \SI{53}{\percent} compared to the best vanilla SNAP gate protocol for realistic target operations. Additionally, our optimization is compatible with the fault-tolerant scheme of \cite{chi-matching-drive,formal-PI,SNAP-ET} that aims to suppress incoherent errors, and both approaches combined yield higher fidelities than each of them individually.

\begin{figure}
	\includegraphics[width=\linewidth]{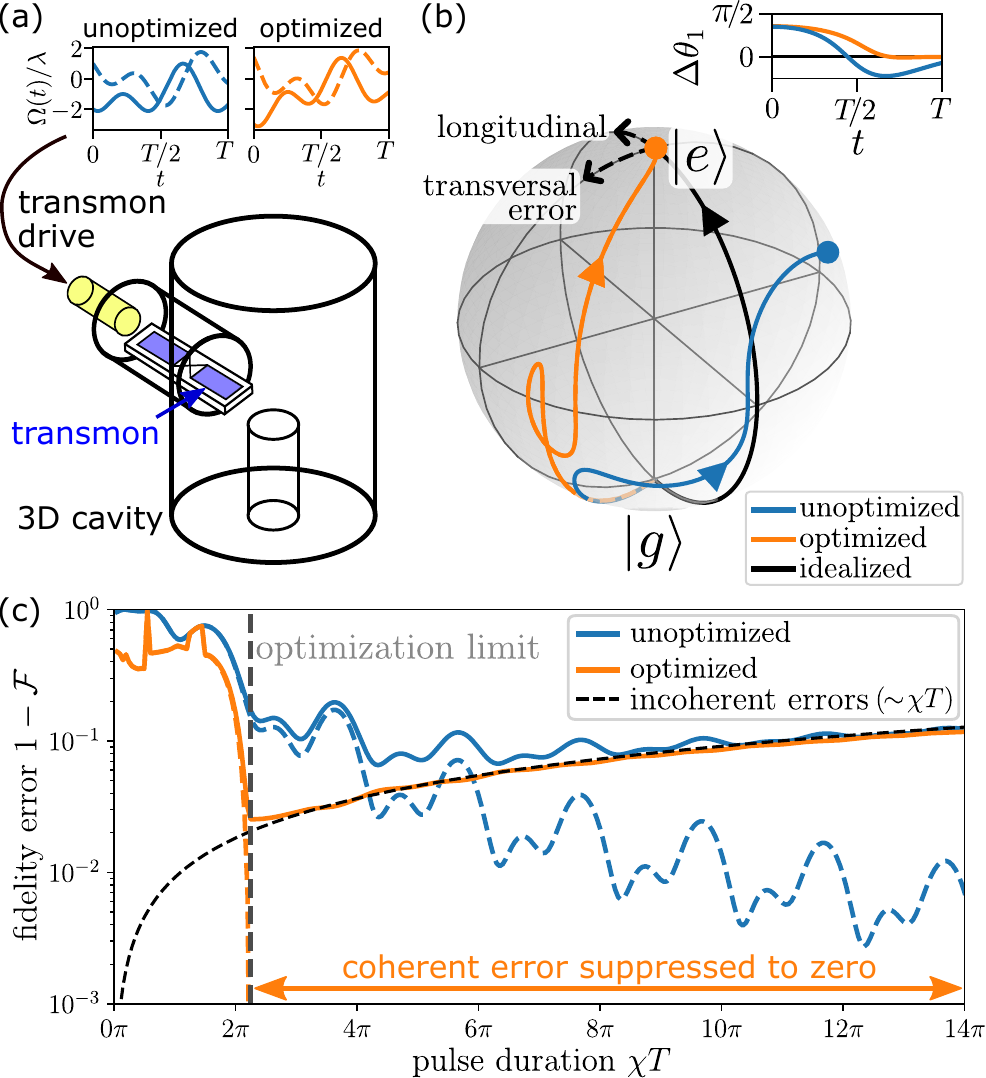}
	\caption{\label{fig:coherent-errors} Experimental setup and principle of our optimization scheme. (a) Circuit QED platform: a transmon (blue) is coupled to a 3D microwave cavity and a readout resonator (yellow). By driving the transmon with a slow pulse $\Omega(t)$ (solid lines: $\Re(\Omega(t))$, dashed lines: $\Im(\Omega(t)))$, the target phase shifts are applied selectively to the Fock states of the cavity, implementing the first stage of the SNAP gate. Our scheme improves (orange) the unoptimized (blue) pulse sequences in \cref{equ:unoptimized_pulse} by adapting the amplitudes, frequencies and phases of the individual drive terms.
	(b) Simulated coherent time evolution for one selected Fock state [here Fock state $n=1$ with $\Delta \theta_{1}(t)=\arg(\bra{e,1}\ket{\psi(t)})-\theta_1$, $\chi T=2.25\pi$, and $\vec{\theta}=(0, -\pi/4, \pi/2)$] for the unoptimized (blue) and optimized (orange) pulses shown in (a) in comparison to the idealized trajectory (black).
    (c)	Theoretical estimation of the SNAP gate fidelity error (solid lines) for $\vec{\theta}=(0, -\pi/4, \pi/2)$. For unoptimized pulses, the coherent error (dashed blue) decreases for larger pulse durations, but the incoherent error (dashed black, analytical approximation, for derivation see \cite{supplement}) increases with the pulse duration $\sim \chi T$. Our scheme reduces the coherent error to zero above the optimization limit (gray). 
}
\end{figure}

We perform our experiments in a cQED setup consisting of a 3D storage cavity which is dispersively coupled to a transmon qubit \cite{koch2007charge} (see \cref{fig:coherent-errors}(a)) to manipulate the state in the cavity. Moreover, the transmon is coupled to a readout resonator to measure the transmon state and to drive the transmon. The readout will not be included in our description of the system. Initially, the transmon is prepared in the ground state, and the cavity can be in an arbitrary state. Our SNAP gate consists of two stages: In the first stage, a slow and selective $\pi$ pulse is applied to excite the transmon and to apply the target phase shifts $\vec{\theta}$ to the respective Fock states. The $\pi$ pulse of the second stage is fast and so unselective and just brings the transmon back into the ground state. Ideally, this implements the SNAP gate:
\begin{equation}
	\label{equ:SNAP_operation}
	\mathrm{SNAP}(\vec{\theta}) = \sum_n \mathrm{e}^{\mathrm{i}\theta_n} \ketbra{n}{n}
\end{equation}
on the cavity.

However, the selectiveness of the first stage is only given in the limit of large pulse durations resulting in incoherent errors, while shorter pulses lead to coherent errors. In contrast, the second stage is almost error-free and will not be addressed by our optimization scheme. In the following, we discuss how the coherent errors of the first stage can be classified and how our optimization scheme deals with them.
In the dispersive coupling limit, the Hamiltonian of the cavity-transmon system can be written as \cite{koch2007charge, krastanov2015universal, heeres2015cavity}:
\begin{equation}
	\label{equ:hamiltonian}
	\hat{H} = \hat{H}_0 + \hat{H}_\chi + \hat{H}_\mathrm{drive}
\end{equation}
with $\hat{H}_0=\omega_c \hat{a}^\dag\hat{a} + \omega_{ge} \ketbra{e}{e}$ as the cavity and transmon energy, $\hat{H}_\chi = - \chi \ketbra{e}{e} \hat{a}^\dag\hat{a}$ as the dispersive coupling between both and $\hat{H}_\mathrm{drive} = \Omega(t) \mathrm{e}^{-\mathrm{i}\omega_{ge} t} \ketbra{e}{g} + \mathrm{H.c.}$ as the transmon drive. $\omega_{ge}$ is the transition frequency between the first two transmon states $\ket{g}$ and $\ket{e}$, $\ket{n}$ a Fock state of the cavity, $\omega_c$ the cavity frequency, $\chi$ the dispersive coupling frequency, $\hat{a}$/$\hat{a}^\dag$ the destruction/creation operator of a cavity excitation and $\Omega(t)$ the pulse envelope applied on the transmon (see Appendix I for the concrete system parameters). For simplicity, we work in the following in the frame rotating with $\hat{H}_0 + \hat{H}_\chi$. Note that for practical applications, also higher order contributions to \cref{equ:hamiltonian} have to be considered (see Appendix E \cite{supplement}).

We expand the input state as $\ket{\psi_\mathrm{in}}=\sum_n c_n \ket{gn}$ with $\vec{c}$ as the normalized and complex amplitude vector. As our Hamiltonian in \cref{equ:hamiltonian} conserves the photon number, the amplitude vector is time independent, while the initial $\ket{gn}$ states evolve in time. During the first stage, the pulse function of the unoptimized SNAP protocol \cite{heeres2015cavity,krastanov2015universal}
\begin{equation}
	\label{equ:unoptimized_pulse}
	\Omega_\mathrm{unopt}(t) = \lambda \sum_{n} \mathrm{e}^{\mathrm{i}(\chi n t+ \theta_n)}
\end{equation}
is applied with $\lambda=\pi/(2T)$ as the unoptimized pulse amplitude and $T$ as the pulse duration. Each of the terms aims to resonantly drive the corresponding transition $\ket{gn}\leftrightarrow\ket{en}$ which holds in the limit of large gate times $\chi T\gg 2\pi$ due to a rotating wave approximation (RWA).
In this limit, for each Fock state the coherent time evolution follows a perfect half circle on the Bloch sphere and the circle's azimutal orientation is determined by the corresponding target phase $\theta_n$. The coherent time evolution reaches the target state $\ket{\psi_\mathrm{target}(\vec{c})} = \sum_n c_n \mathrm{e}^{\mathrm{i}\theta_n}\ket{en}$ (see \cref{fig:coherent-errors}(b)) \cite{supplement}.
However, for finite pulse durations terms neglected in the RWA have to be taken into account. Thus, the actual trajectories differ and the final states deviate for each Fock state from the target ones in the following three ways: (i) the acquired phases are shifted by the phase errors $\Delta \theta_n$, (ii) on the Bloch sphere the trajectories overshoot or stop too early, denoted as the longitudinal errors $\epsilon_n^{(L)}$, and (iii) the end points on the Bloch sphere deviate perpendicular to the orientation of the ideal trajectory, labeled as the transversal errors $\epsilon_n^{(T)}$. In general, the actual final state $\ket{\psi_\mathrm{out}(\vec{c})}$ is given with these three error contributions as:

\begin{equation}
    \label{equ:actual_terminal_state}
	\sum_n c_n \Bigg( \sqrt{1-\frac{\abs{\epsilon_n}^2}{4}} \mathrm{e}^{\mathrm{i}(\theta_n + \Delta \theta_n)}\ket{en} - \frac{\epsilon_n}{2} \ket{gn} \Bigg)
\end{equation}

with $\epsilon_n=(\epsilon_n^{(L)} + \mathrm{i} \epsilon_n ^{(T)})\mathrm{e}^{\mathrm{i}\Delta \theta_n}$.


To compensate these errors, we modify the pulse sequence of \cref{equ:unoptimized_pulse} in the following way:
\begin{equation}
	\label{equ:optimized_pulse}
	\Omega_\mathrm{opt}(t) = \sum_n \lambda_n \mathrm{e}^{\mathrm{i}(\omega_n t + \alpha_n - \Delta \omega_n T /2)}
\end{equation}
with $\lambda_n$, $\omega_n$ and $\alpha_n$ as the Fock level dependent amplitudes, frequencies and phases of the drive.  $\Delta \omega_n = \omega_n - \chi n$ is the detuning with respect to the unoptimized frequencies of \cref{equ:unoptimized_pulse}. The phase $-\Delta \omega_n T/2$ corrects an unwanted phase shift created by the detuning. 
Using first order perturbation theory in the limit of large gate times and small coherent errors, the latter can be corrected by modifying the pulse parameters by (see Appendix C):
\begin{align}
	\label{equ:correction_paras}
	\Delta \lambda_n = - \frac{\epsilon_n^{(L)}}{2T};  \quad
	\Delta \alpha_n = - \Delta \theta_n; \quad
    \Delta \omega_n = \frac{\pi \epsilon_n^{(T)}}{2T}
\end{align}
Thus, each of the coherent errors is controlled individually by one pulse parameter.

However, for finite gate times and correspondingly large coherent errors, higher order contributions to \cref{equ:correction_paras} become important and the updated pulses still lead to coherent errors. To overcome this problem, we iteratively reapply \cref{equ:correction_paras}. First, we simulate the coherent time evolution with the current pulse parameters and extract the coherent errors. Next, we update the pulse parameters according to \cref{equ:correction_paras}. To prevent overshooting, the updates are scaled by a learning rate $\eta \in (0,1]$. This routine is repeated until the convergence fails or the coherent mean overlap error is smaller than a threshold, typically $10^{-5}$, where it is experimentally negligible. Therefore, our scheme is able to correct coherent errors far beyond the range of validity of the first order perturbation theory. As shown in \cref{fig:coherent-errors}(b), the time evolution under the optimized pulse still deviates from the idealized trajectory for intermediate times. However, the desired target state is reached nevertheless.


In \cref{fig:coherent-errors}(c), the overall performance of the SNAP gate is shown. The coherent errors of the unoptimized SNAP gate decay with increasing gate time (scaling with $1/(\chi T)^2$, see \cite{supplement}), while the incoherent errors rise with the gate time (scaling with $\chi T$). Thus, the unoptimized SNAP gate is ideally operated at intermediate gate times. In contrast, our optimized pulses completely suppress coherent errors provided that the gate time exceeds a certain threshold, which we denote the "optimiziation limit". This limit depends on the actual target operation. Thus, our optimized SNAP gates reach their best performance at the optimization limit. 

\begin{figure*}
	\includegraphics[width=\linewidth]{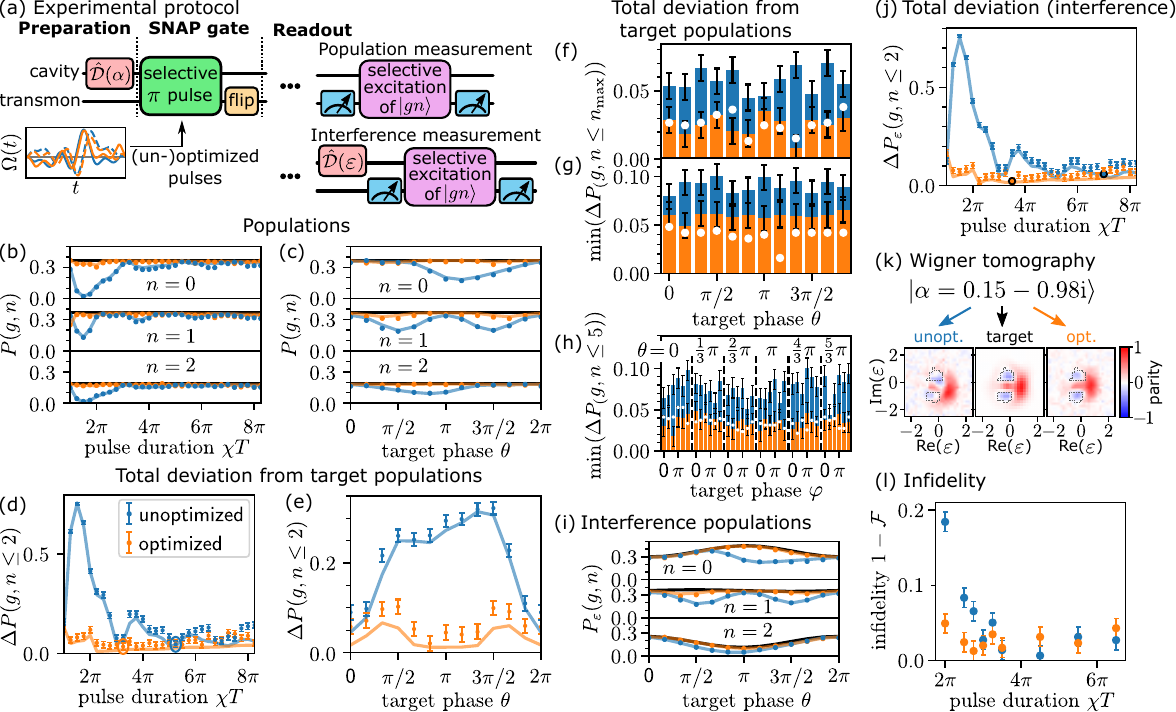}
	\caption{\label{fig:excited_state_error}
	Experimental results, showing performance improvements by our optimized SNAP protocol. (a) Experimental procedure: state preparation, SNAP gate, and readout. We measure the populations $P(g,n)$ (top right), or extract the coherences via an interference measurement (bottom right) yielding $P_\epsilon(g,n)$ after a small displacement $\epsilon$. Measurement symbols indicate a dispersive measurement of the qubit. (b,c) Photon-number-dependent ground state population $P(g,n)$ as function of the pulse duration (b) and target operation phase $\theta$ (c). The unoptimized (blue) and optimized (orange) pulses (dots: experiment) are compared to the ideal populations $P_\mathrm{ideal} (g,n)$ (black) for an error-free SNAP gate (black). Simulation results (solid) are obtained from a Lindblad master equation (see \cite{supplement}). The statistical measurement error is smaller than the size of the points. (d, e) Total deviation $\Delta P(g, n\leq n_\mathrm{max})$ between the ideal and the observed populations. The encircled points in (d) mark the best performance of the unoptimized and optimized SNAP gate. (f--h) Smallest total deviation with respect to the gate time for various target operations. White points indicate the theoretically expected residual error for the optimised pulse. (i, j) Interference measurement populations with $\epsilon=0.1$ and their total deviations from their ideal values. (k) Wigner tomography of the cavity state after the unoptimised (fidelity: \SI{82\pm 1}{\percent}) and optimised (fidelity: \SI{95\pm 1}{\percent}) protocol was applied, compared to the desired target state. The contour line of the target state for $\mathrm{parity} = 0$ is shown in gray in the inner region of all 3 insets. (l) Infidelity of the final cavity state estimated from the Wigner tomographies.\\
	Considered target operations: $(0,\pi,0)$ in (b, d, j), $(0,\theta,0)$ in (c, e, f, i), $(0,\theta,0,\theta,0,\theta)$ in (g), $(0,\varphi,\theta,0,\pi,0)$ in (h), and $(0,\frac{2}{3}\pi, \pi, 0, \pi, 0)$ in (k,l). Considered gate times: $\chi T =2.5\pi$ in (c, e, i), and $\chi T=2\pi$ in (k). Considered initial coherent states: $\alpha=1.0$ in (b--f, h--j), $\alpha=1.4$ in (g), and $\alpha=0.15-0.98\mathrm{i}$ in (k,l).
    }
\end{figure*}

To showcase the potential of our approach in the experiment, we first prepare the cavity in a coherent state of amplitude $\alpha$ (see \cref{fig:excited_state_error}(a)), using a displacement operation $\mathcal{\hat{D}}$. Then we apply the unoptimzed/optimized SNAP gate.

To determine the quality of the gate, we perform two different kinds of measurements on the final state. First, we determine the populations of the cavity-transmon system $P(g,n)$. To that end, we first measure the transmon, obtaining $P(g)$ and $P(e)$. Afterwards, the transition $\ket{gn}\leftrightarrow\ket{en}$ for a chosen Fock state $n$ is driven selectively. A subsequent measurement of the transmon provides the Fock state occupancies $P(n|g/e)$ conditioned on the first readout result. The resulting populations $P(g,n)$ gives information about the combined longitudinal and transversal errors defined in \cref{equ:actual_terminal_state}. If noise is neglected, $P(g,n)$ equals $\abs{c_n}^2 (1-\abs{\epsilon_n}^2/4)$. Furthermore, one can distinguish the longitudinal and transversal error by measuring the operators $\hat{\sigma}_{x/y}\otimes \ketbra{n}{n}$ (see Appendix M) which is beyond the scope of this work.


Second, to get information about the phase errors, we apply a small displacement $\epsilon$ to the final cavity state and measure again the populations of the cavity-transmon system, labeled as $P_\varepsilon(g,n)$. Due to the small displacement, neighboring Fock state components interfere and the resulting populations depend on the phase shifts acquired during the SNAP gate operation \cite{heeres2015cavity}. We choose $\epsilon=0.1$, which leads to a strong interference between neighboring modes (see \cref{fig:excited_state_error}(i)), but is still small enough to prevent interference beyond directly neighboring Fock modes. 


In \cref{fig:excited_state_error}(b), our optimized pulses are compared against the standard SNAP gate. As intended, our optimization scheme minimizes individually for each Fock state the deviation from the ideal ground state population. Comparing the smallest total deviations $\Delta P(g, n\leq 2)=\sum_{n=0}^2 |P_\mathrm{ideal} (g,n)-P(g,n)|$ (see \cref{fig:excited_state_error}(d)), our approach lowers the total excited state population error from \num{0.046} to \num{0.035}, while shortening the dimensionless pulse time $\chi T$ from \num{5.25}$\pi$ to \num{3.25}$\pi$.

The size of the improvement depends on the target operation. Some target  operations, like $\vec{\theta}=(0,0,0)$ (see \cref{fig:excited_state_error}(c,e)) have small coherent errors, here shown for a fixed pulse duration of $\chi T=2.5\pi$, leaving little room for improvement. In contrast, other target operations, like $\vec{\theta}=(0,\pi,0)$ lead to large coherent errors and our approach clearly outperforms the standard SNAP gate. In \cref{fig:excited_state_error}(f--h), we compare the smallest achieved total excited state errors for a variety of target operations. For a more complex example, as shown in (h), our scheme decreases the deviation $\Delta P(g, n\leq 5)$ from \num{0.076} to \num{0.035} (a reduction by \SI{53}{\percent}) and the dimensionless gate time $\chi T$ from \num{5.58}$\pi$ to \num{4.44}$\pi$ (a reduction by \SI{42}{\percent}) when averaged over the target phase parameters. 

As shown in \cref{fig:excited_state_error}(i,j) our scheme also strongly reduces the deviations of the interference measurement compared to the ideal evolution and always outperforms the unoptimized protocol. In this measurement, the deviations originate both from excited state errors and phase errors. An upper bound for the phase error can be estimated at $\SI{0.24}{\radian}$ (see \cite{supplement}), which is unfortunately insufficient to resolve the improvement predicted by the simulations.



The ability of our scheme to successfully prevent errors is also shown by the Wigner tomography of the final cavity state in \cref{fig:excited_state_error}(k). We estimate the SNAP gates' fidelities from the Wigner tomographies, see \cref{fig:excited_state_error}(l), and our optimised SNAP gates reach overall significantly smaller fidelities for short gate times.

\begin{figure}
	\includegraphics[width=\linewidth]{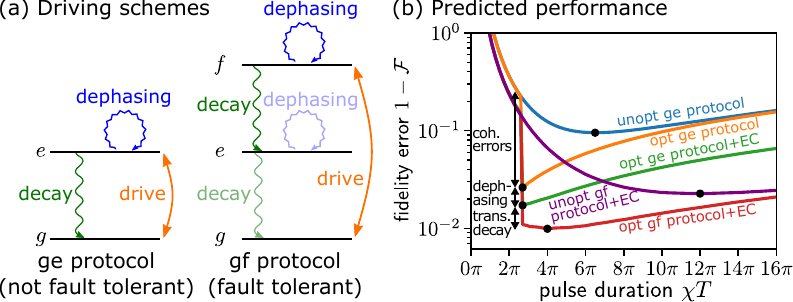}
	\caption{\label{fig:error_contributions} Combination of our optimization with the fault-tolerant scheme of \cite{chi-matching-drive,formal-PI,SNAP-ET}. (a) Comparison of the $ge$ driving scheme and the $gf$ driving scheme and their dominant transmon noise channels (dominant error channels: high saturation, second order processes: low saturation). (b) Theoretically predicted performance. The fidelity error is averaged over all possible target operations and the oscillations in $\chi T$ for different SNAP protocols (best performances highlighted by black dots): unoptimized $ge$ protocol without error correction (EC) (blue), optimized $ge$ protocol without EC (orange), optimized $ge$ protocol with EC (green), unoptimized $gf$ protocol with EC (purple) and optimized $gf$ protocol with EC (red). The fidelity errors are estimated by perturbation theory (see \cite{supplement} for more details).}
\end{figure}

While our approach focuses on the suppression of the coherent errors, the fault-tolerant scheme presented in \cite{chi-matching-drive,formal-PI,SNAP-ET} aims to reduce the incoherent errors from transmon decay and transmon dephasing. The key idea is to make the SNAP gate path-independent, i.e.~that the final cavity-transmon state is independent of when, how many and which quantum jumps have occurred. This is achieved by including the second excited transmon state $f$ and making use of feedback. 

In \cref{fig:error_contributions}(b), we estimate theoretically how the fidelity error scales for different SNAP protocols for realistic parameter values. The SNAP gate is applied to the first four Fock states. The error is averaged over all initial quantum states, all target operations, and over the oscillations in $\chi T$ (see \cref{fig:coherent-errors}(c)). As discussed above, the unoptimized $ge$ (blue) is operated best at intermediate gate times, while our scheme (orange) performs best at the optimization limit. The $ge$ SNAP protocol is in the limit of large gate times path-independent with respect to transmon decay errors \cite{formal-PI}. By measuring the final transmon state transmon dephasing errors can be detected and corrected (green). Due to violations of the path-independence for finite gate times, a small error is remaining, which is discussed further in \cite{supplement}. We obtain the best results by combining our approach with the $gf$ driving scheme (red), which is in the limit of large gate times path-independent with respect to transmon dephasing and decay \cite{chi-matching-drive,formal-PI,SNAP-ET}. The remaining errors are cavity decay and the path-independence violations for transmon dephasing and decay for finite gate times. 


In conclusion, our optimized SNAP gates completely suppress coherent errors, given that the gate time exceeds a certain threshold, and, in combination with the fault-tolerant scheme in \cite{chi-matching-drive,formal-PI,SNAP-ET}, can further reduce incoherent errors. Furthermore, using our optimized SNAP gates together with displacement operations opens the way to efficiently implement arbitrary unitary operations \cite{krastanov2015universal,fosel2020efficient, kudra2022robust}. While the optimization in this work was performed in simulations, we foresee the possibility of optimizing the pulse shapes directly based on experimental data, employing a feedback loop where pulse parameters are adapted suitably (see Appendix M \cite{supplement}). Furthermore, we anticipate that our scheme opens the road to optimise other gates which rely on selectiveness, like applying Fock-state selective rotations or photon additions (called the SNAPPA gate \cite{kudra2022experimental}). Extending our scheme to the SNAPPA gates could find applications in autonomous quantum error correction and the realisation of arbitrary non-unitary operations. Beyond the field of cQED, we foresee that our scheme could be of used to correct incoherent errors in an array of frequency-multiplexed qubits all driven by the same global pulse \cite{shi2023multiplexed,takeuchi2024microwave}.


\begin{acknowledgments}
We thank James Teoh, Takahiro Tsunoda, William Kalfus, Sal Elder, Hugo Ribeiro, and especially Jacob Curtis for fruitful discussions. The research is part of the Munich Quantum Valley, which is supported by the Bavarian state government with funds from the Hightech Agenda Bayern Plus. R.J.S. was supported by the U.S. Army Research Office (ARO) under grant W911NF-18-1-0212. The views and conclusions contained in this document are those of the authors and should not be interpreted as representing official policies, either expressed or implied, of the ARO or the U.S. Government. The U.S. Government is authorized to reproduce and distribute reprints for Government purpose notwithstanding any copyright notation herein.
\end{acknowledgments}

\section*{Competing interests}
R.J.S. is a founder and shareholder of Quantum Circuits, Inc. The remaining authors declare no competing interests.

\bibliography{main.bib}



\end{document}


\title{Supplemental Material for ``Fast quantum control of cavities using an improved protocol without coherent errors''}
\maketitle

\onecolumngrid
\begin{appendix}

\section{Ideal SNAP gate dynamics for $T\gg 1/\chi$}
\label{supp:ideal_dynamics}
With the system Hamiltonian (see main text) and an arbitrary drive function $\Omega (t)$, the Hamiltonian in the frame rotating with $\hat{H}_0+\hat{H}_\chi$ turns out to equal:
\begin{align}
    \label{equ:SNAP_Hamiltonian_rot}
    \hat{H}_\mathrm{rot} = &\sum_n \left( \hat{\sigma}_x \Re \left(  \Omega(t) \mathrm{e}^{-\mathrm{i}\chi n t}  \right) +  \hat{\sigma}_y \Im \left(  \Omega(t) \mathrm{e}^{-\mathrm{i}\chi n t}  \right)  \right) \ketbra{n}{n}
\end{align}
with $\hat{\sigma}_i$ ($i=\{x,y,z\}$) as the Pauli operators. For the unoptimized drive (see Eq.~(3) in the main text) and performing a rotating wave approximation valid in the limit $T\gg 2\pi/\chi$, the Hamiltonian simplifies to:
\begin{equation}
	\label{equ:approx_Hamiltonian}
	\hat{H}_\mathrm{rot} \approx \lambda \sum_{n}  \hat{\sigma}_{\alpha_n} \ketbra{n}{n}
\end{equation}
with $\hat{\sigma}_{\alpha_n}=\hat{\sigma}_x \cos \alpha_n + \hat{\sigma}_y \sin \alpha_n$ as the $\hat{\sigma}_x$ Pauli operator rotated by the angle $\alpha_n$ around the positive $z$-axis. With the transmon intialized in the ground state $\ket{g}$ and the cavity in an arbitrary initial state $\sum_n c_n \ket{n}$, the time evolution of this state turns out to equal:
\begin{equation}
	\label{equ:total_state_ideal_evolution}
	\ket{\psi(t)} = \sum_n c_n \left( \cos (\lambda t) \ket{gn} -\mathrm{i} \sin (\lambda t) \mathrm{e}^{\mathrm{i}\alpha_n} \ket{en} \right)
\end{equation}
Setting the pulse amplitude to $\lambda=\pi/(2T)$, the target state of the first stage of the SNAP gate is reached:
\begin{equation}
    \ket{\psi}_\mathrm{target}=\sum_n c_n \mathrm{e}^{\mathrm{i}\theta_n}\ket{en}
\end{equation}
with the phase shifts $\theta_n = \alpha_n-\pi/2$.

Of further use is also the time evolution operator of the ideal dynamics:
\begin{equation}
    \label{equ:time_evolution_operator_ideal}
    \hat{U}_\mathrm{ideal}(t) = \mathrm{e}^{-\mathrm{i}\hat{H}_{\mathrm{rot}} t} =\sum_n \left[\hat{\mathds{1}} \cos \left(\lambda t  \right) - \mathrm{i} \hat{\sigma}_{\alpha_n} \sin \left(\lambda t  \right)\right] \otimes \ketbra{n}{n}
\end{equation}
At the end time $T$ the time evolution operator equals:
\begin{equation}
    \hat{U}_\mathrm{ideal}(T) =  \sum_n -\mathrm{i} \hat{\sigma}_{\alpha_n} \otimes \ketbra{n}{n}
\end{equation}

\section{Graphical interpretation of the coherent errors}
\label{supp:graphical_interpretation}
\begin{figure}
	\centering
	\includegraphics[width=0.3\columnwidth]{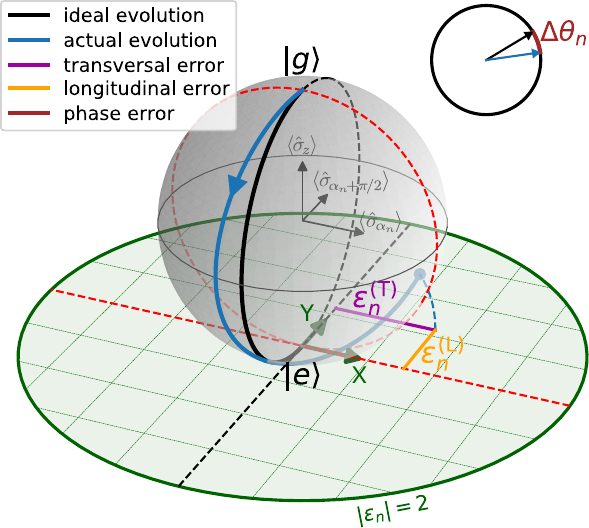}
	\caption{Geometrical definition of the coherent errors on the Bloch sphere. The final state of each Fock level dependent transmon state is projected onto the Lambert azimuthal equal-area projection plane (green). 
	The elongation of the ideal trajectory in the projection plane ($Y$-direction, dashed black) determines the direction of the longitudinal error $\varepsilon_n^{(L)}$, its perpendicular line ($X$-direction, dashed red) the direction of the transversal error $\varepsilon_n^{(T)}$. The phase error $\Delta \theta_n$ is visualized in the phasor representation.}
	\label{fig:coherent_error_definition}
\end{figure}

In this section, we show the connection between the quantum state error definition in Eq.~(4) and our geometrical interpretation. In the Bloch sphere coordinate system ($\langle \hat{\sigma}_{\alpha_n} \rangle$,$\langle \hat{\sigma}_{\alpha_n+\pi/2} \rangle$,$\langle \hat{\sigma}_z \rangle$), the ideal trajectory corresponds to a perfect half circle around the $(1,\ 0,\ 0)$ axis. The terminal state after the first stage of the SNAP gate (see Eq.~(4) in the main text) has the coordinates:
\begin{equation}
    \label{equ:terminal_state_coordinates}
    \begin{pmatrix}
        \langle \hat{\sigma}_{\alpha_n} \rangle\\
        \langle \hat{\sigma}_{\alpha_n+\pi/2} \rangle\\
        \langle \hat{\sigma}_z \rangle
    \end{pmatrix}=
	\begin{pmatrix}
		\sqrt{1-\frac{1}{4}\abs{\varepsilon_n}^2}\epsilon_n^{(T)}\\
		\sqrt{1-\frac{1}{4}\abs{\varepsilon_n}^2}\epsilon_n^{(L)}\\ %
		\frac{1}{2}\abs{\varepsilon_n}^2 - 1
	\end{pmatrix}
\end{equation}
To identify a graphical interpretation, we use the Lambert azimuthal equal-area projection at the $\ket{e}$ point:
\begin{equation}
	\label{equ:Lambert-projection}
	(X,\ Y) = \left(\sqrt{\frac{2}{1-\langle \hat{\sigma}_z \rangle}}\langle \hat{\sigma}_{\alpha_n} \rangle,\ \sqrt{\frac{2}{1-\langle \hat{\sigma}_z \rangle}}\langle \hat{\sigma}_{\alpha_n +\pi/2} \rangle \right)
\end{equation}
The coordinates X and Y in the projection plane, turn out to equal $X=\epsilon_n^{(T)}$ and $Y=\epsilon_n^{(L)}$ for the terminal state. Thus, the transversal error corresponds to a deviation perpendicular to the ideal trajectory, while the longitudinal error corresponds to an over-/undershoot in direction of the ideal trajectory in the projection plane.

\section{Derivation of the optimization scheme}
\label{supp:deriv_opt_scheme}
In this section, we investigate how our correction parameters of our optimized driving protocol (see Eq.~(5) in the main text) influence the coherent errors and can thus be used to correct these errors. Compared to the unoptimized driving scheme (see Eq.~(3) in the main text), our drive has three Fock-level-dependent correction parameters: the amplitude correction $\Delta \lambda_n = \lambda_n - \lambda $ with $\lambda=\pi/(2T)$, the frequency correction $\Delta \omega_n=\omega_n - \chi n$, and the phase correction $\Delta \alpha_n = \alpha_n - \theta_n - \pi/2$. \\

\noindent\textbf{Amplitude correction}\quad We replace $\lambda$ in \cref{equ:total_state_ideal_evolution} by $\lambda_n = \pi/2T+\Delta\lambda_n$ and expand for small $\Delta \lambda_n$:
\begin{equation}
	\ket{\psi(T)} \approx \sum_n c_n \left( -\Delta \lambda_n T \ket{gn}+\mathrm{e}^{\mathrm{i}\theta_n}\ket{en} \right)
\end{equation}
Thus, amplitude corrections $\Delta \lambda_n$ create longitudinal errors $\varepsilon_n^{(L)} \approx \pi \Delta\lambda_n/\lambda$ on their associated Fock states $n$, while the transversal and phase errors remain approximately zero. 

Performing the same replacement in \cref{equ:time_evolution_operator_ideal}, we can furthermore approximate the time evolution operator as:
\begin{equation}
    \hat{U}(T) \approx \hat{U}_\mathrm{ideal}(T) - \sum_n \Delta \lambda_n T \hat{\mathds{1}} \otimes \ketbra{n}{n}
\end{equation}

\noindent\textbf{Phase correction}\quad Substituting $\alpha_n$ in \cref{equ:total_state_ideal_evolution} by $\theta_n+\pi/2 + \Delta \alpha_n$, the final state turns out to equal:
\begin{equation}
	\ket{\psi(T)} = \sum_n c_n \mathrm{e}^{\mathrm{i}(\theta_n+\Delta\alpha_n)} \ket{en}
\end{equation}
Therefore, the amplitude shift $\Delta \alpha_n$ creates a phase error $\Delta \theta_n=\alpha_n$ and the other coherent errors remain zero.

The time evolution operator can be approximated as:
\begin{equation}
    \hat{U}(T) \approx \hat{U}_\mathrm{ideal}(T) - \mathrm{i} \sum_n \Delta \alpha_n \hat{\sigma}_{\alpha_n + \pi/2} \otimes \ketbra{n}{n}
\end{equation}

\noindent\textbf{Detuning}\quad The driving frequencies $\omega_n$ for each Fock state $n$ are now detuned by $\Delta \omega_n$ from the unoptmized drive frequencies $\chi n$ with $\Delta \omega_n t \ll 1$:
\begin{align}
    \label{equ:pulse_detuning}
	\Omega(t) &= \sum_n \lambda \mathrm{e}^{\mathrm{i}(\chi n t + \Delta \omega_n t + \alpha_n)} \overset{\Delta \omega_n t \ll 1}{\approx} \sum_n \lambda \left(1 + \mathrm{i}\Delta \omega_n t\right)\mathrm{e}^{\mathrm{i}(\chi n t + \alpha_n)}
\end{align}
By inserting this pulse into \cref{equ:SNAP_Hamiltonian_rot} and taking only terms close to resonance into account, our Hamiltonian reads as:
\begin{equation}
	\hat{H}_{\mathrm{rot}} \approx \sum_n \left( \lambda \hat{\sigma}_{\alpha_n} + \lambda \Delta\omega_n t \hat{\sigma}_{\alpha_n+\pi/2}\right) \ketbra{n}{n}
\end{equation}
The first summand in this Hamiltonian corresponds to the ideal dynamics (see also \cref{supp:ideal_dynamics}), the second term results from the detuning. 

We now switch into the frame moving along with the ideal trajectory:
\begin{align}
    \begin{split}
	\hat{\widetilde{H}}_{\mathrm{rot}} \approx \sum_n \lambda \Delta \omega_n t & \left( \hat{\sigma}_{\alpha_n+\pi/2} \cos(2\lambda t)- \hat{\sigma}_z \sin(2\lambda t)  \right) \ketbra{n}{n}
	\end{split}
\end{align}
which leads us to the time evolution operator in this frame:
\begin{align}
	\hat{\widetilde{U}}(T) &\approx \hat{\mathds{1}} - i\int_0^T \hat{\widetilde{H}}_{\mathrm{rot}} (t)\,dt = \hat{\mathds{1}} + \sum_n \mathrm{i}\Delta \omega_n T \left(\frac{1}{\pi}\hat{\sigma}_{\alpha_n+\pi/2} + \frac{1}{2}\hat{\sigma}_z\right)\ketbra{n}{n}
\end{align}
Back in the frame rotating with $\hat{H}_0 + \hat{H}_\chi$, the time evolution operator equals:
\begin{equation}
    \hat{U}(T) = \hat{U}_\mathrm{ideal}(T) + \sum_n\left[\frac{\mathrm{i}\Delta \omega_n T}{\pi}\hat{\sigma}_z - \frac{\mathrm{i} \Delta\omega_n T}{2} \hat{\sigma}_{\alpha_n+\pi/2}\right]\ketbra{n}{n}
\end{equation}

Thus, the final state back in the original frame turns out to equal:
\begin{equation}
	\ket{\psi (T)} \approx \sum_n c_n \left(\frac{\mathrm{i}}{\pi}\Delta\omega_n T \ket{gn} + \mathrm{e}^{\mathrm{i}(\theta_n+\Delta\omega_n T/2)} \ket{en}\right)
\end{equation}
with a vanishing longitudinal error, a transversal error $\varepsilon_n^{(T)} \approx -\Delta\omega_n/\lambda$ and a phase error $\theta_n \approx \Delta\omega_n T/2 $. To compensate for the unwanted phase shift, the drive phases are modified in Eq.~(5) by $-\Delta\omega_n T/2$ and thus we can freely control the transversal error without disturbing the phase.\newline

\noindent\textbf{Summary for all correction parameters}\quad As demonstrated, each of the coherent errors can be individually controlled by one of the correction parameters of our optimized pulse shape. Even though our derivation is performed in the limit of large pulse durations $T$ and small correction parameters, iteratively reapplying Eq.~(6) leads to optimized pulse sequences far beyond the validity range of our approximations. \newline

\noindent\textbf{Intuitive explanation}\quad The effect of each pulse parameter on the coherent errors can be understood very intuitively: (i) Increasing the drive amplitude leads to a faster rotation on the Bloch sphere and, for a fixed gate time, to an overshoot error $\epsilon_n^{(L)}$. (ii) Modifying the drive phase changes the acquired phase accordingly. (iii) A small detuning slowly rotates the circle's axis on which the trajectory is moving resulting in a transversal deviation.\newline

\noindent\textbf{Impact on the SNAP gate operator}\quad Our modified pulse sequence (see Eq.~(3)) applies the gate:
\begin{equation}
    \hat{U}(T) \approx \hat{U}_\mathrm{ideal}(T) + \sum_n\left[-\Delta \lambda_n T \hat{\mathds{1}} + \frac{\mathrm{i}\Delta \omega_n T}{\pi}\hat{\sigma}_z - \mathrm{i} \Delta \alpha_n \hat{\sigma}_{\alpha_n + \pi/2}  \right]\ketbra{n}{n}
\end{equation}
Therefore, amplitude corrections add identity contributions, detunings lead to $\hat{\sigma}_z$ contributions, and phase corrections lead to $\hat{\sigma}_{\alpha_n + \pi/2}$ contributions.

\section{Fidelity definition}

To quantify the overall performance of the first SNAP pulse, we define the fidelity as the mean squared overlap:
\begin{equation}
    \label{equ:mean_squared_error_no_noise_ge_protocol}
    \mathcal{F}=\underset{\substack{\vec{c} \text{ s.t.}\\ \norm{c}=1}}{\mathrm{avg}} \bra{\psi_\mathrm{target}(\vec{c})} \hat{\rho}_\mathrm{out}(\vec{c}) \ket{\psi_\mathrm{target}(\vec{c})}
\end{equation}
where the average covers all possible initial cavity states. $\hat{\rho}_\mathrm{out}(\vec{c})$ is the density matrix of the cavity-transmon system after the first stage of the SNAP gate and equals $\ket{\psi_\mathrm{out}(\vec{c})}\bra{\psi_\mathrm{out}(\vec{c})}$ if only the coherent dynamics are considered.

\section{Kerr nonlinearity and correction to the dispersive coupling}
\label{supp:kerr}
For the application in the actual experiment also higher order contributions to the Hamiltonian in Eq.~(2) have to be considered. This includes the Kerr nonlinearity $\hat{H}_\mathrm{Kerr}=-(K/2) \hat{a}^{\dag^2} \hat{a}^2$ and the correction to the dispersive coupling $\hat{H}_{\chi^\prime}=(\chi^\prime/2) \ketbra{e}{e} \hat{a}^{\dag^2} \hat{a}^2$. To understand their effect on the dynamics of the SNAP gate, we transform the Hamiltonian into the frame rotating with $\hat{H}_0+\hat{H}_\chi+\hat{H}_\mathrm{Kerr}+\hat{H}_{\chi^\prime}$:
\begin{align}
	\begin{split}
	\hat{H}_{\mathrm{rot}} &=
	\sum_n   \left[\hat{\sigma}_x \Re\left(\Omega(t)\mathrm{e}^{-\mathrm{i}\left(\chi n - \frac{1}{2}\chi^\prime(n^2-n)\right)t}\right) + \hat{\sigma}_y \Im\left(\Omega(t)\mathrm{e}^{-\mathrm{i}\left(\chi n - \frac{1}{2}\chi^\prime(n^2-n)\right)t}\right) \right] \ketbra{n}{n}
	\end{split}
\end{align}
So, to drive the $\ket{gn}\leftrightarrow\ket{en}$ resonantly, we replace the driving frequencies by $\omega_n=\chi n - \chi^\prime(n^2-n)/2$. Using the rotating wave approximation, the Hamiltonian is again identical with \cref{equ:approx_Hamiltonian}. However, the target operation is defined in the system rotating with $\hat{H}_0+\hat{H}_\chi$. This change in frame has to be compensated by setting the unoptimized drive phases $\alpha_n$ to $\theta_n + \pi/2 - (K-\chi^\prime)(n^2-n)T/2$. In the limit of large pulse durations, and neglecting noise, this implements again the first stage of the SNAP gate.

\section{Full system dynamics including the second excited transmon state and noise}
\label{supp:full_system_dynamics}

\subsection{System Hamiltonian including the second excited transmon state}

Including the second excited transmon state $\ket{f}$, we redefine the components of our system Hamiltonian $\hat{H} = \hat{H}_0 + \hat{H}_\chi + \hat{H}_\mathrm{drive}$ as:
\begin{subequations}
	\begin{align}
		\hat{H}_0 &= \omega_{ge}\ketbra{e}{e} + \omega_{gf}\ketbra{f}{f}  \\
		\hat{H}_\chi &= - \chi \ketbra{e}{e}\hat{a}^\dag\hat{a} - \chi_f \ketbra{f}{f} \hat{a}^\dag\hat{a}
	\end{align}
\end{subequations}
with $\omega_{gf}$ as the transition frequency between the ground and second excited transmon state $\ket{g}$ and $\ket{f}$, and $\chi_f$ as the dispersive coupling frequency with the $\ket{f}$ state. To make the gf SNAP protocol fault-tolerant with respect to transmon decay, we set $\chi_f=\chi$ to fulfill the $\chi$-matching condition \cite{chi-matching-drive,formal-PI,SNAP-ET} in our analysis. For the ge drive protocol, the drive Hamiltonian $\hat{H}_\mathrm{drive}$ is defined in the main text, for the gf protocol by $\hat{H}_\mathrm{drive}=\Omega(t) \mathrm{e}^{-\mathrm{i}\omega_{gf} t} \ketbra{f}{g} + \mathrm{H.c.}$. For simplicity, we neglect the Kerr nonlinearity and the correction to the dispersive coupling in the comparison between the different driving protocols.

In the frame comoving with $\hat{H}_0+\hat{H}_\chi$, the Hamiltonian for the gf SNAP protocol equals:
\begin{equation}
    \label{equ:H_rot_gf_SNAP}
    \hat{H}_\mathrm{rot} = \sum_n \left( (\ketbra{f}{g}+\ketbra{g}{f}) \Re \left(  \Omega(t) \mathrm{e}^{-\mathrm{i}\chi n t}  \right) +  \mathrm{i}(\ketbra{f}{g}-\ketbra{g}{f}) \Im \left(  \Omega(t) \mathrm{e}^{-\mathrm{i}\chi n t}  \right)  \right)\otimes \ketbra{n}{n}
\end{equation}
Using the rotating wave approximation and the unoptimized drive (see Eq.~(3) in the main text), this simplifies to:
\begin{equation}
    \label{equ:approx_Hamiltonian_gf_SNAP}
    \hat{H}_\mathrm{rot} \approx \lambda \sum_n \left(\exp(\mathrm{i}\alpha_n)\ketbra{f}{g}+\exp(-\mathrm{i}\alpha_n)\ketbra{g}{f}\right)\otimes\ketbra{n}{n}
\end{equation}
Therefore, the time evolution for an initial state $\ket{\psi (0)}=\sum_n c_n \ket{gn}$ turns out to equal (compare to \cref{equ:total_state_ideal_evolution}):
\begin{equation}
	\label{equ:total_state_ideal_evolution_f}
	\ket{\psi(t)} = \sum_n c_n \left( \cos (\lambda t) \ket{gn} -\mathrm{i} \sin (\lambda t) \mathrm{e}^{\mathrm{i}\alpha_n} \ket{fn} \right)
\end{equation}
For $\lambda=\pi/(2T)$, the target state of the first stage of the SNAP gate is reached:
\begin{equation}
    \ket{\psi}_\mathrm{target}=\sum_n c_n \mathrm{e}^{\mathrm{i}\theta_n}\ket{fn}
\end{equation}

\subsection{Non-unitary time evolution}
\label{supp:Lindblad_master_equation}
Above, we developed a scheme to completely suppress the coherent errors. We considered noise only indirectly as we also minimized the gate time. However, to maximize the gate performance we have to optimize the total error including noise. The dominant noise channels are transmon decay, transmon dephasing and cavity decay. The corresponding Lindblad operators in the frame comoving with $\hat{H}_0+\hat{H}_\chi$ are given by \cite{SNAP-ET}:
\begin{subequations}
    \begin{align}
        \hat{L}_{e\rightarrow g} &= \mathrm{e}^{\mathrm{i}\chi \hat{a}^\dag \hat{a}t} \ketbra{g}{e} \\
        \label{equ:lindblad_operator_f_e_decay}
        \hat{L}_{f\rightarrow e} &= \mathrm{e}^{\mathrm{i}(\chi_f-\chi) \hat{a}^\dag \hat{a}t} \ketbra{e}{f} \\
        \label{equ:lindblad_ee_dephasing}
        \hat{L}_{ee} &= \ketbra{e}{e}\\
        \hat{L}_{ff} &= \ketbra{f}{f}\\
        \label{equ:lindblad_cavity_decay}
        \hat{L}_\mathrm{cav} &= \mathrm{e}^{\mathrm{i}(\chi \ketbra{e}{e}+\chi_f \ketbra{f}{f})t} \hat{a}
    \end{align}
\end{subequations}
with $L_{e\rightarrow g}$ and $L_{f\rightarrow e}$ as the transmon decay operators, $L_{ee}$ and $L_{ff}$ as the transmon dephasing operators, and $L_\mathrm{cav}$ as the cavity decay operator.


The time evolution of the density matrix $\hat{\rho}$ under the influence of noise is described by the Lindblad Master equation:
\begin{equation}
	\label{equ:Lindblad_Master_equ}
	\frac{\mathrm{d}\hat{\rho}}{\mathrm{d}t} = -\mathrm{i}\left[\hat{H}_\mathrm{rot}, \hat{\rho}\right] + \sum_j \Gamma_j \left( \hat{L}_j \hat{\rho} \hat{L}_j^\dag - \frac{1}{2}\left\{\hat{\rho}, \hat{L}_j^\dag \hat{L}_j \right\} \right)
\end{equation}
with $\hat{H}_\mathrm{rot}$ as the Hamiltonian in the rotating frame comoving with $\hat{H}_0+\hat{H}_\chi$. The sum iterates over all noise contributions with $\Gamma_j$ as the corresponding noise rate.

\subsection{Propagation of errors during the ge SNAP gate and definition of the mean squared overlap}
\label{supp:error_propagation_ge_SNAP}
Here, we analyze following \cite{SNAP-ET}, how transmon decay and dephasing errors are propagated during the ge SNAP gate operation in the limit of large $\chi T$, where the system Hamiltonian is given by \cref{equ:approx_Hamiltonian}. Initially, the system is in the state $\ket{\psi_\mathrm{in}}=\sum_n c_n \ket{gn}$. The time evolution consists of three steps: First, the the dynamics follow the idealized system Hamiltonian from \cref{equ:approx_Hamiltonian} up to the time $t=t_j$, where a quantum jump event (either transmon decay or dephasing) occurs. After the jump, the evolution follows again the idealized Hamiltonian up to $t=T$.

For a single transmon dephasing event at $t_j$, the final quantum state turns out to equal:
\begin{subequations}
	\label{equ:PI_dephasing_ee}
	\begin{align}
		\ket{\psi(T)} &\propto \mathrm{e}^{-\mathrm{i}\hat{H}_\mathrm{rot} (T-t_j)} \hat{L}_{ee} \mathrm{e}^{-\mathrm{i}\hat{H}_\mathrm{rot} t_j} \ket{\psi_\mathrm{in}}\\
		&\propto - \sin (\lambda(T-t_j))\ket{\psi_\mathrm{in}} + \cos(\lambda (T-t_j)) \ket{\psi_\mathrm{target}} 
	\end{align}
\end{subequations}
Directly after the first stage of the SNAP gate was performed, we measure the qubit state. If the outcome is $\ket{e}$, the final state equals the target state $\ket{\psi_\mathrm{target}}=\sum_n c_n \mathrm{e}^{\mathrm{i}\theta_n}\ket{en}$. If we measure $\ket{g}$, the state is the initial state $\ket{\psi_\mathrm{in}}$, as if no pulse would have been applied. To correct this error, one has to reapply the SNAP pulse. These results are independent of the actual jump time and so the $ge$ SNAP protocol is fault tolerant for at least one dephasing event. Based on this result, we define the target states for a measurement outcome $\ket{g}$ and $\ket{e}$ as:
\begin{subequations}
    \label{equ:target_states_ge_protocol}
    \begin{align}
        \ket{\psi_{\mathrm{target},g}} &= \ket{\psi_\mathrm{in}} =\sum_n c_n \ket{gn} \\
        \ket{\psi_{\mathrm{target},e}} &= \ket{\psi_\mathrm{target}} = \sum_n c_n \mathrm{e}^{\mathrm{i}\theta_n}\ket{en}
    \end{align}    
\end{subequations}
Repeating the same analysis for transmon decay $e\rightarrow g$, results in:
\begin{subequations}
	\label{equ:PI_decay_ge}
	\begin{align}
		\ket{\psi(T)} &\propto \mathrm{e}^{-\mathrm{i}\hat{H}_\mathrm{rot} (T-t_j)} \hat{L}_{e\rightarrow g} \mathrm{e}^{-\mathrm{i}\hat{H}_\mathrm{rot} t_j} \ket{\psi(0)}\\
		&\propto \sum_{n\in\mathcal{N}} c_n\mathrm{e}^{\mathrm{i}(\chi n t_j+\theta_n)} \left[ \cos(\lambda (T-t_j)) \ket{gn} + \sin(\lambda (T-t_j))\mathrm{e}^{\mathrm{i}\theta_n} \ket{en} \right]
	\end{align}
\end{subequations}
As the final state after an $\ket{g}$ and $\ket{e}$ measurement depends on the jump time $t_j$, the ge SNAP protocol is not fault tolerant with respect to transmon decay.

Based on the target states in \cref{equ:target_states_ge_protocol}, we define the the mean squared overlap for ge protocol without error correction as (see \cref{equ:mean_squared_error_no_noise_ge_protocol}):
\begin{equation}
    \mathcal{F}=\underset{\substack{\vec{c} \text{ s.t.}\\ \norm{c}=1}}{\mathrm{avg}} \bra{\psi_{\mathrm{target},e}(\vec{c})} \hat{\rho}_\mathrm{out}(\vec{c}) \ket{\psi_{\mathrm{target},e}(\vec{c})}
\end{equation}
and for the ge protocol with error correction:
\begin{equation}
    \mathcal{F}=\underset{\substack{\vec{c} \text{ s.t.}\\ \norm{c}=1}}{\mathrm{avg}} \sum_{j=\{g,e\}} \bra{\psi_{\mathrm{target},j}(\vec{c})} \hat{\rho}_\mathrm{out}(\vec{c}) \ket{\psi_{\mathrm{target},j}(\vec{c})}
\end{equation}
with $\hat{\rho}_\mathrm{out}(\vec{c})$ as the density matrix of the cavity-transmon system after the first stage of the SNAP gate.

\subsection{Propagation of errors during the gf SNAP gate and definition of the mean squared overlap}
\label{supp:error_propagation_gf_SNAP}
In this section, we repeat the procedure of \cref{supp:error_propagation_ge_SNAP} for the gf SNAP protocol in the limit of large gate times, where the Hamiltonian is approximated by \cref{equ:approx_Hamiltonian_gf_SNAP}. The two dominant noise channels are $f\rightarrow e$ decay and transmon dephasing. As the ideal SNAP evolution now ends in the $f$ state, we redefine the target state as $\ket{\psi_\mathrm{target}}=\sum_n c_n \mathrm{e}^{\mathrm{i}\theta_n}\ket{fn}$. If a dephasing event occurs at time $t=t_j$, the final system state is given by:
\begin{subequations}
	\begin{align}
		\ket{\psi(T)} &\propto \mathrm{e}^{-\mathrm{i}\hat{H}_\mathrm{rot} (T-t_j)} \hat{L}_{ff} \mathrm{e}^{-\mathrm{i}\hat{H}_\mathrm{rot} t_j} \ket{\psi_\mathrm{in}}\\
		&\propto - \sin (\lambda(T-t_j)) \ket{\psi_\mathrm{in}} + \cos(\lambda (T-t_j)) \ket{\psi_\mathrm{target}}
	\end{align}
\end{subequations}
In case of an $f\rightarrow e$ decay event, the final state is:
\begin{subequations}
	\begin{align}
		\ket{\psi(T)} &\propto \mathrm{e}^{-\mathrm{i}\hat{H}_\mathrm{rot} (T-t_j)} \hat{L}_{f\rightarrow e} \mathrm{e}^{-\mathrm{i}\hat{H}_\mathrm{rot} t_j} \ket{\psi_\mathrm{in}}\\
		&\propto \sum_{n\in\mathcal{N}} c_n \mathrm{e}^{\mathrm{i}\left((\chi_f-\chi)nt_j +\theta_n\right)} \ket{en} \overset{\chi=\chi_f}{=} \sum_n c_n \mathrm{e}^{\mathrm{i}\theta_n}\ket{en}
	\end{align}
\end{subequations}
Thus, the measurement result $\ket{g}$ corresponds to a dephasing error and the system ends up in its initial state. The measurement result $\ket{e}$ corresponds to a transmon decay error, where the cavity state was correctly prepared, but the transmon ends up in $\ket{e}$. This can be corrected by simply flipping the transmon state. In case of an $\ket{f}$ measurement, the SNAP gate was successfully applied. Correspondingly, we define the target states for the different measurement outcomes as:
\begin{subequations}
    \label{equ:target_states_gf_protocol}
    \begin{align}
        \ket{\psi_{\mathrm{target},g}} &= \ket{\psi_\mathrm{in}} =\sum_n c_n \ket{gn} \\
        \label{equ:target_state_e}
        \ket{\psi_{\mathrm{target},e}} & = \sum_n c_n \mathrm{e}^{\mathrm{i}\theta_n}\ket{en} \\
        \ket{\psi_{\mathrm{target},f}} &= \ket{\psi_\mathrm{target}} = \sum_n c_n \mathrm{e}^{\mathrm{i}\theta_n}\ket{fn}
    \end{align}    
\end{subequations}
Therefore, we define the mean squared overlap for the gf SNAP gate with error correction as:
\begin{equation}
    \mathcal{F}=\underset{\substack{\vec{c} \text{ s.t.}\\ \norm{c}=1}}{\mathrm{avg}} \sum_{j=\{g,e,f\}} \bra{\psi_{\mathrm{target},j}(\vec{c})} \hat{\rho}_\mathrm{out}(\vec{c}) \ket{\psi_{\mathrm{target},j}(\vec{c})}
\end{equation}

\section{Estimation of the dominant error contributions}
\label{supp:dominant_errors}
In the following sections, we derive analytically the dominant error contributions for the various SNAP gate implementations mentioned in the main text. The dominant errors include the coherent errors, transmon decay, transmon dephasing, cavity decay and path-independence violations. The errors are always averaged over all initial cavity states, defined by the amplitude vector $\vec{c}$, all possible target operations, given by $\vec{\theta}$, and the oscillations in $\chi T$.

\subsection{Coherent errors}
\label{supp:coherent_errors_scaling}
Here, we derive the coherent error contributions in the limit of large $\chi T$ for the unoptimized pulse from Eq.~(3) and the scaling law of the mean-squared-overlap error. The notation throughout this section is for the $ge$ SNAP protocol, but the result is the same for the $gf$ protocol.

Using the unoptimized drive (see Eq.~(3) in the main text), we rewrite the Hamiltonian from \cref{equ:SNAP_Hamiltonian_rot} as:
\begin{equation}
	\hat{H}_{\mathrm{rot}} = \sum_n \sum_{m} \lambda \hat{\sigma}_{(\chi t (m-n)+\alpha_m)} \ketbra{n}{n}
\end{equation}
with $\hat{\sigma}_\alpha=\hat{\sigma}_x \cos \alpha + \hat{\sigma}_y \sin \alpha $ as a Pauli operator rotated by $\alpha$ around the z axis. We transform this Hamiltonian into the frame rotating with the idealized SNAP Hamiltonian from \cref{equ:approx_Hamiltonian}:
\begin{align}
\begin{split}
	\hat{\widetilde{H}}_{\mathrm{rot}}
	&= \sum_n  \sum_{m\neq n} \lambda \left[ \cos^2(\lambda t) \hat{\sigma}_{(\chi t (m-n)+\alpha_m)} + \sin^2(\lambda t) \hat{\sigma}_{(2\alpha_n - \alpha_m -\chi t(m-n))} + \right. \\ 
	& \qquad \qquad \left. + \sin(2\lambda t) \sin (\alpha_n - \alpha_m -\chi t(m-n))\hat{\sigma}_z \right] \ketbra{n}{n}
	\end{split}
\end{align}
The time evolution operator in this rotating frame is given by:

\begin{align}
	\hat{\widetilde{U}} &= \hat{\mathds{1}} - \mathrm{i} \int_0^T\,dt\, \hat{\widetilde{H}}_{\mathrm{rot}}(t) - \int_0^T \,dt \int_0^t \,dt^\prime \hat{\widetilde{H}}_{\mathrm{rot}}(t) \hat{\widetilde{H}}_{\mathrm{rot}}(t^\prime) + \mathrm{...}
\end{align}

Solving the integrals, sorting in powers of $1/(\chi T)$ and applying the time evolution operator to the initial state $\ket{gn}$, results in:
\begin{equation}
	\hat{\widetilde{U}} \ket{gn} = \ket{gn} + \sum_{m\neq n} \frac{\pi}{2(m-n)\chi T}  \left[ \mathrm{e}^{\mathrm{i}((n-m)\chi T + 2\alpha_n - \alpha_m)} +\mathrm{e}^{\mathrm{i}\alpha_m} - \mathrm{e}^{\mathrm{i}\alpha_n} \right] \ket{en} + \mathcal{O}((\chi T)^{-2})
\end{equation}
By transforming the result back into the frame rotating with $\hat{H}_0 + \hat{H}_\chi$, we receive the final state $|\psi_\mathrm{out}\rangle$:
\begin{align}
	|\psi_\mathrm{out}\rangle = \sum_n \Big( \mathrm{e}^{\mathrm{i}\theta_n} \ket{en} + \sum_{m\neq n} \frac{-\mathrm{i}\pi}{2(m-n)\chi T} \big( \mathrm{e}^{\mathrm{i}((n-m)\chi T + \theta_n - \theta_m)} +\mathrm{e}^{\mathrm{i}(\theta_m-\theta_n)} - 1 \big) \ket{gn} \Big) + \mathcal{O}((\chi T)^{-2})
\end{align}
with the coherent error contributions (see Eq.~(4) in the main text):
\begin{subequations}
	\label{equ:scaling_coherent_errors}
	\begin{align}
		\varepsilon_n &= \sum_{m\neq n} \frac{\mathrm{i}\pi}{(m-n)\chi T} \left[ \mathrm{e}^{\mathrm{i}((n-m)\chi T + \theta_n - \theta_m)} +\mathrm{e}^{\mathrm{i}(\theta_m-\theta_n)} - 1 \right] + \mathcal{O}((\chi T)^{-2}) \\
		\Delta \theta_n &= \mathcal{O}((\chi T)^{-2}) 
	\end{align}
\end{subequations}
$\varepsilon_n$, and so also the longitudinal and transversal errors, scale with $1/(\chi T)$. Our calculations show that the phase error $\Delta \theta_n$ scales at least with $1/(\chi T)^2$. Our simulations confirm that $1/(\chi T)^2$ is the correct scaling law.

The mean squared overlaps $\mathcal{F}_g$ and $\mathcal{F}_e$ with the target states $\ket{\psi_{\mathrm{target},g}}$ and $\ket{\psi_{\mathrm{target},e}}$ turn out to equal:
\begin{align}
	\mathcal{F}_g &= \underset{\substack{\vec{c} \in \mathbb{C}^L \text{ s.t.}\\ \norm{c}=1}}{\mathrm{avg}} \abs{\langle \psi_{\mathrm{target},g} | \psi_{\mathrm{out}}\rangle}^2 = \underset{\substack{\vec{c} \in \mathbb{C}^L \text{ s.t.}\\ \norm{c}=1}}{\mathrm{avg}}\frac{1}{4}\abs{\sum_n \abs{c_n}^2 \epsilon_n}^2 =\frac{1}{4}\sum_{n,n^\prime} \overbrace{\underset{\substack{\vec{c} \in \mathbb{C}^L \text{ s.t.}\\ \norm{c}=1}}{\mathrm{avg}}(\abs{c_n c_{n^\prime}}^2) }^{=(1+\delta_{n,n^\prime})/(L(L+1))} \epsilon_{n^\prime}^\ast \epsilon_n		= \sum_{n,n^\prime}\frac{1+\delta_{n,n^\prime}}{4L(L+1)}\epsilon_{n^\prime}^\ast \epsilon_n
\end{align}
\begin{subequations}
	\begin{align}
		\mathcal{F}_e &= \underset{\substack{\vec{c} \in \mathbb{C}^L \text{ s.t.}\\ \norm{c}=1}}{\mathrm{avg}} \abs{\langle \psi_{\mathrm{target},e} | \psi_{\mathrm{out}}\rangle}^2 = \underset{\substack{\vec{c} \in \mathbb{C}^L \text{ s.t.}\\ \norm{c}=1}}{\mathrm{avg}} \abs{\sum_n \abs{c_n}^2 \sqrt{1-\abs{\epsilon_n}^2}\mathrm{e}^{\mathrm{i}\Delta\theta_n}}^2\\
		&=\sum_{n,n^\prime} \overbrace{\underset{\substack{\vec{c} \in \mathbb{C}^L \text{ s.t.}\\ \norm{c}=1}}{\mathrm{avg}}(\abs{c_n c_{n^\prime}}^2) }^{=(1+\delta_{n,n^\prime})/(L(L+1))}  \left[ \sqrt{(1-\abs{\varepsilon_n}^2/4)(1-\abs{\varepsilon_{n^\prime}}^2/4)} \mathrm{e}^{\mathrm{i} (\Delta \theta_n-\Delta \theta_{n^\prime}) }  \right] = \\
		&= 1- \frac{1}{L}\sum_n \frac{\abs{\epsilon_n}^2}{4} + \mathcal{O}((\chi T)^{-3})
	\end{align}
\end{subequations}
with $L$ as the number of Fock states part of the target operation.

Next, we derive the squared mean overlap error averaged over the target phases and $\chi T$ oscillations. Using \cref{equ:scaling_coherent_errors}, we get the relation:
\begin{equation}
	\underset{\vec{\theta},\ \chi T \mathrm{ osc.}}{\mathrm{avg}} \varepsilon_{n^\prime}^\ast \varepsilon_n = \left(\frac{\pi}{\chi T}\right)^2 \sum_{\substack{m,m^\prime \\ m \neq n \\ m^\prime \neq n^\prime}} \frac{1}{(m-n)(m^\prime-n^\prime)} \left[ 2\delta_{n,n^\prime} \delta_{m,m^\prime} + 1 \right] + \mathcal{O}((\chi T)^{-3})
\end{equation}

Thus, the averaged mean squared overlaps equal:
\begin{align}
    \underset{\vec{\theta},\ \chi T \mathrm{ osc.}}{\mathrm{avg}} \mathcal{F}_g = \frac{5}{4L(L+1)}\left(\frac{\pi}{\chi T}\right)^2\sum_{\substack{n,m \\ m \neq n}} \frac{1}{(m-n)^2} + \mathcal{O}((\chi T)^{-3})
\end{align}
\begin{align}
    \underset{\vec{\theta},\ \chi T \mathrm{ osc.}}{\mathrm{avg}} \mathcal{F}_e = 1-\frac{3}{4L}\left(\frac{\pi}{\chi T}\right)^2\sum_{\substack{n,m \\ m \neq n}} \frac{1}{(m-n)^2} + \mathcal{O}((\chi T)^{-3})
\end{align}
For all unoptimized SNAP protocols in Fig.~3 without error correction, the coherent error contribution is estimated by $1-\underset{\vec{\theta},\ \chi T \mathrm{ osc.}}{\mathrm{avg}}\mathcal{F}_e$, for all unoptmized protocols with error correction, by $1-\underset{\vec{\theta},\ \chi T \mathrm{ osc.}}{\mathrm{avg}}\mathcal{F}_g-\underset{\vec{\theta},\ \chi T \mathrm{ osc.}}{\mathrm{avg}}\mathcal{F}_e$.

\subsection{$e\rightarrow g$ decay}
\label{supp:eg-decay}
In this section, we derive the transmon decay error for the $ge$ SNAP gate in the limit of $\chi T \to \infty$, but $\Gamma_{e\rightarrow g} T \to 0$. All other noise rates are set to zero throughout this section. We first expand the density matrix in orders of $\Gamma_{e\rightarrow g} T$:
\begin{equation}
	\hat{\rho}(t) = \hat{\rho}_0 (t) + \hat{\rho}_1 (t) + ...
\end{equation}
where $\hat{\rho}_j(t)$ is of order $(\Gamma_{e\rightarrow g} T)^j$. With this expansion of the density matrix and the idealized Hamiltonian from \cref{equ:approx_Hamiltonian}, we sort the Lindblad Master equation from \cref{equ:Lindblad_Master_equ} in orders of $\Gamma_{e\rightarrow g} T$:
\begin{subequations}
	\begin{align}
		\derivative{\hat{\rho}_0(t)}{t} &= -\mathrm{i} [\lambda\sum_n \hat{\sigma}_{\alpha_n} \ketbra{n}{n}, \hat{\rho}_0(t)] \\
		\label{equ:LBME_expansion_eg_decay_rho1}
		\derivative{\hat{\rho}_1(t)}{t} &= -\mathrm{i} [\lambda\sum_n \hat{\sigma}_{\alpha_n} \ketbra{n}{n}, \hat{\rho}_1(t)] + \Gamma_{e\rightarrow g} \left[\hat{L}_{e\rightarrow g} \hat{\rho}_0 (t) \hat{L}^\dag_{e\rightarrow g} - \frac{1}{2} \left\{\hat{\rho}_0(t), \hat{L}^\dag_{e\rightarrow g} \hat{L}_{e\rightarrow g} \right\} \right] \\
		&\shortvdotswithin{=} \notag
	\end{align}
\end{subequations}
The solution for the unperturbed density matrix $\hat{\rho}_0$ is simply given by the coherent evolution:
\begin{equation}
	\label{equ:solution_eg_decay_rho0}
	\hat{\rho}_0 (t) = \ketbra{\psi(t)}{\psi(t)}
\end{equation}
with the idealized time evolution of the quantum state $\ket{\psi(t)}$, defined in \cref{equ:total_state_ideal_evolution}. To solve \cref{equ:LBME_expansion_eg_decay_rho1}, we go into the frame comoving with the ideal trajectory. With the ideal time evolution operator $\hat{U}(t)=\mathrm{exp}\left(-\mathrm{i}\sum_n \lambda \hat{\sigma}_{\alpha_n} \ketbra{n}{n} t\right)$ and $\hat{\widetilde{\rho}}_1(t) = \hat{U}(t)^\dag \hat{\rho}_1(t)  \hat{U}(t)$, \cref{equ:LBME_expansion_eg_decay_rho1} can be rewritten as:
\begin{subequations}
	\begin{align}
		\derivative{\hat{\widetilde{\rho}}_1(t)}{t} &= \Gamma_{e\rightarrow g} \hat{U}(t)^\dag \left[\hat{L}_{e\rightarrow g} \hat{\rho}_0 (t) \hat{L}^\dag_{e\rightarrow g} - \frac{1}{2} \left\{\hat{\rho}_0(t), \hat{L}^\dag_{e\rightarrow g} \hat{L}_{e\rightarrow g} \right\} \right] \hat{U}(t) \\
		\begin{split}
			\label{equ:eg_decay_step}
			&= \Gamma_{e\rightarrow g} \big[ \sum_{n,n^\prime} c_n c_{n^\prime}^\ast  \overbrace{\mathrm{e}^{\mathrm{i}\chi (n-n^\prime)t}}^{\xrightarrow[]{\substack{\text{rotating}\\ \text{wave approx.}}} \delta_{n,n^\prime}}\mathrm{e}^{\mathrm{i}(\theta_n - \theta_{n^\prime})} \sin^2 (\lambda t) \hat{U}^\dag(t) |gn\rangle\langle gn^\prime | \hat{U}(t)  \\
			&\quad - \frac{1}{2} \hat{U}^\dag(t) \left\{\ketbra{\psi(t)}{\psi(t)}, \ketbra{e}{e} \right\} \hat{U}(t)\big]
		\end{split}
	\end{align}
\end{subequations} 

Using the rotating wave approximation, we ignore all fast oscillating terms. 
With $\hat{\widetilde{\rho}}_1(0)=0$, $\hat{\widetilde{\rho}}_1(T)$ is obtained by integrating \cref{equ:eg_decay_step} in time. Transforming the result back into the original frame, we receive:
\begin{align}
	\begin{split}
		\hat{\rho}_1 (T) = \Gamma_{e\rightarrow g} T \left[ \sum_n \abs{c_n}^2 
		\begin{pmatrix}
			\frac{3}{8} & \frac{1}{2\pi}\mathrm{e}^{-\mathrm{i}\theta_n}\\
			\frac{1}{2\pi}\mathrm{e}^{\mathrm{i}\theta_n} & \frac{1}{8}
		\end{pmatrix} 
		\otimes \ketbra{n}{n} \right.
		\left.-\frac{1}{2}\sum_{n,n^\prime} c_n c_{n^\prime}^\ast 
		\begin{pmatrix}
			0 & -\frac{1}{\pi}\mathrm{e}^{-\mathrm{i}\theta_{n^\prime}} \\
			-\frac{1}{\pi}\mathrm{e}^{\mathrm{i}\theta_n} & \mathrm{e}^{\mathrm{i}(\theta_n - \theta_{n^\prime})}
		\end{pmatrix}
		\otimes \ketbra{n}{n^\prime} \right] 
	\end{split}
\end{align}
Thus, we get the mean squared overlaps:
\begin{equation}
	\mathcal{F}_g = \frac{3}{4(L+1)} \Gamma_{e\rightarrow g}T + \mathcal{O}((\Gamma_{e\rightarrow g}T)^2,(\Gamma_{e\rightarrow g}T)(\chi T)^{-1})
\end{equation}
\begin{equation}
	\mathcal{F}_e = 1-\frac{2L+1}{4(L+1)} \Gamma_{e\rightarrow g}T +\mathcal{O}((\Gamma_{e\rightarrow g}T)^2,(\Gamma_{e\rightarrow g}T)(\chi T)^{-1})
\end{equation}

The ge SNAP protocol without error correction has an error contribution of $1-\mathcal{F}_e$, the ge SNAP protocol with error correction has an error of $1-\mathcal{F}_g-\mathcal{F}_e$. In contrast, the gf SNAP protocol with error correction is fault tolerant with respect to the dominant transmon decay channel, namely the $f\rightarrow e$ decay, in the limit of large $\chi T$. Therefore, it does not suffer from transmon decay in first order. However, for finite $\chi T$ the path-independence is violated. The resulting errors are further discussed in \cref{supp:ET_violations}.

\subsection{Transmon dephasing}
\label{supp:transmon_dephasing}
In analogy to \cref{supp:eg-decay}, we derive in this section the effect of transmon dephasing on the ge SNAP protocol in the limit of $\chi T \to \infty$ and $\Gamma_{ee} T\to 0$. Expanding the density matrix $\hat{\rho}(t)=\hat{\rho}_0(t)+\hat{\rho}_1(t)+...$ in orders of $\Gamma_{ee}T$ and ordering the Lindblad Master equation in orders of $\Gamma_{ee}T$, results in:
\begin{subequations}
	\begin{align}
		\label{equ:LBME_expansion_ee_dephasing_rho0}
		\derivative{\hat{\rho}_0(t)}{t} &= -\mathrm{i} [\lambda\sum_n \hat{\sigma}_{\alpha_n} \ketbra{n}{n}, \hat{\rho}_0(t)] \\
		\label{equ:LBME_expansion_ee_dephasing_rho1}
		\derivative{\hat{\rho}_1(t)}{t} &= -\mathrm{i} [\lambda\sum_n \hat{\sigma}_{\alpha_n} \ketbra{n}{n}, \hat{\rho}_1(t)] + \Gamma_{ee} \left[\hat{L}_{ee} \hat{\rho}_0 (t) \hat{L}^\dag_{ee} - \frac{1}{2} \left\{\hat{\rho}_0(t), \hat{L}^\dag_{ee} \hat{L}_{ee} \right\} \right] \\
		&\shortvdotswithin{=} \notag
	\end{align}
\end{subequations}
with $\hat{L}_{ee}$ as the Lindblad operator for dephasing, see \cref{equ:lindblad_ee_dephasing}. The solution for the unperturbed density matrix $\hat{\rho}_0$ is given by \cref{equ:solution_eg_decay_rho0}. As in \cref{supp:eg-decay}, we solve \cref{equ:LBME_expansion_ee_dephasing_rho1} by transforming it into the frame comoving with the ideal trajectory and integrate it in time, with $\hat{\rho}_1(0)=0$. The final state $\hat{\rho}_1(T)$ turns out to equal:
\begin{equation}
    \hat{\rho}_1(T)=\frac{1}{8}\Gamma_{ee}T\sum_{n,n^\prime} c_n c_{n^\prime}^\ast \left( \ketbra{g}{g} - \mathrm{e}^{\mathrm{i}(\theta_n-\theta_{n^\prime})}\ketbra{e}{e}\right)\otimes \ketbra{n}{n^\prime}
\end{equation}
and results in the mean squared overlaps:
\begin{equation}
    \mathcal{F}_g = \frac{1}{8}\Gamma_{ee}T + \mathcal{O}((\Gamma_{ee}T)^2,(\Gamma_{ee}T)(\chi T)^{-1})
\end{equation}
\begin{equation}
    \label{equ:fid_e_ee_dephasing}
    \mathcal{F}_e = 1-\frac{1}{8}\Gamma_{ee}T + \mathcal{O}((\Gamma_{ee}T)^2,(\Gamma_{ee}T)(\chi T)^{-1})
\end{equation}
Therefore, the ge SNAP protocol without error correction suffers from transmon dephasing and has a fidelity error of $1-\mathcal{F}_e$. In contrast, the ge SNAP protocol with error correction has a fidelity of $\mathcal{F}_g+\mathcal{F}_e=1$ and is fault tolerant with respect to transmon dephasing. The gf SNAP protocol with error correction is fault tolerant with respect to transmon dephasing, too \cite{formal-PI}.

\subsection{Cavity decay}
\label{supp:cavity-decay}
We assume in this section, that the Fock states $0$ to $L-1$ are part of the target operation. Analogous to section \ref{supp:eg-decay}, we only consider cavity decay and ignore other noise contributions. $\chi T$ is assumed to be large, while $\Gamma_\mathrm{cav} T$ is small. Throughout this section, we use the notation of the ge SNAP protocol. The results are identical with the gf SNAP protocol. 

We expand $\hat{\rho}$ in orders of $\Gamma_\mathrm{cav}T$:
\begin{subequations}
	\begin{align}
		\label{equ:LBME_expansion_cav_decay_rho0}
		\derivative{\hat{\rho}_0(t)}{t} &= -\mathrm{i} [\lambda\sum_n \hat{\sigma}_{\alpha_n} \ketbra{n}{n}, \hat{\rho}_0(t)] \\
		\label{equ:LBME_expansion_cav_decay_rho1}
		\derivative{\hat{\rho}_1(t)}{t} &= -\mathrm{i} [\lambda\sum_n \hat{\sigma}_{\alpha_n} \ketbra{n}{n}, \hat{\rho}_1(t)] + \Gamma_\mathrm{cav} \left[\hat{L}_\mathrm{cav} \hat{\rho}_0 (t) \hat{L}^\dag_\mathrm{cav} - \frac{1}{2} \left\{\hat{\rho}_0(t), \hat{L}^\dag_\mathrm{cav} \hat{L}_\mathrm{cav} \right\} \right] \\
		&\shortvdotswithin{=} \notag
	\end{align}
\end{subequations}
with $\hat{L}_\mathrm{cav}$ from \cref{equ:lindblad_cavity_decay} as the Lindblad operator for cavity decay in the frame rotating with $\hat{H}_0$. The solution for the unperturbed density matrix $\hat{\rho}_0$ is again given by \cref{equ:solution_eg_decay_rho0}. We transform \cref{equ:LBME_expansion_cav_decay_rho1} into the frame comoving with the ideal trajectory:
\begin{align}
	\label{equ:cav_decay_rho1_rot_frame}
	\derivative{\hat{\widetilde{\rho}}_1(t)}{t} = \Gamma_\mathrm{cav} \Big( \hat{U}(t)^\dag \hat{L}_\mathrm{cav} \ketbra{\psi (t)}{\psi (t)} L_\mathrm{cav}^\dag \hat{U}(t) - \frac{1}{2} \sum_{n,n^\prime=0}^{L-1} c_n c_{n^\prime}^\ast (n+n^\prime) |gn\rangle\langle gn^\prime | \Big)
\end{align}
Integrating \cref{equ:cav_decay_rho1_rot_frame} in time and transforming the result back into the frame rotating with $\hat{H}_0+\hat{H}_\chi$ leads us to the final density matrix and the mean squared overlaps:
\begin{equation}
	\mathcal{F}_g = \frac{L-1}{8(L+1)} \Gamma_\mathrm{cav}T + \mathcal{O}((\Gamma_\mathrm{cav}T)^2,(\Gamma_\mathrm{cav}T)(\chi T)^{-1})
\end{equation}
\begin{equation}
	\mathcal{F}_e = 1-\frac{(L-1)(4L+1)}{8(L+1)} \Gamma_\mathrm{cav}T + \mathcal{O}((\Gamma_\mathrm{cav}T)^2,(\Gamma_\mathrm{cav}T)(\chi T)^{-1})
\end{equation}
All SNAP protocols without error correction suffer a fidelity error of $1-\mathcal{F}_e$, all protocols with error correction suffer an error of $1-\mathcal{F}_g-\mathcal{F}_g$. Note again, that this result is only valid, if the Fock states $0$ to $L-1$ are driven. By using only every second cavity mode, like in the binomial code \cite{michael2016new}, cavity decay errors could be detected by parity measurements and either tracked or corrected.

\subsection{Path-independence violations}
\label{supp:ET_violations}

The $ge$ SNAP is only fault tolerant with respect to transmon dephasing in the limit of large $\chi T$. The same is true for the $gf$ SNAP protocol with respect to transmon dephasing and transmon decay \cite{chi-matching-drive,formal-PI,SNAP-ET} (see also \cref{supp:error_propagation_ge_SNAP,supp:error_propagation_gf_SNAP}). For large $\chi T$, either of the initial $\ket{gn}$ states is moving with the same speed along the Bloch sphere, always forming a perfect half circle from $\ket{g}$ to $\ket{e}$, where the direction is defined by $\theta_n$. Therefore, all of the different Fock states always have the same latitude on the Bloch sphere, and the differences between the acquired phase shifts are constant as time evolution takes place. In contrast, the different Fock states move with different speeds for finite gate times, and the phase shifts are not constant. Therefore, the final quantum state will depend on when a quantum jump has occurred, which violates path independence. Furthermore, the pulses optimized with our scheme do not lead to path independence, as our approach only optimizes the endpoint of the evolution and not intermediate points in time. The resulting additional errors are discussed in the following sections. We could not derive a closed, analytical solution for the errors. Instead, we connect the path-dependence errors and the coherent time evolution for finite gate times.

\subsubsection{Path-independence violations for transmon decay for the $gf$ SNAP protocol}
\label{supp:path_independence_violations_decay}

To further quantify the path-independence violations, we define the coherent evolution of the quantum state (or \enquote{no-jump} trajectory) for all times as:
\begin{equation}
	\label{equ:no-jump-trajectory}
	\ket{\psi(t)} = \sum_{n\in\mathcal{N}}c_n \left(  \sqrt{1-\mu_n(t)} \mathrm{e}^{\mathrm{i}\varphi_{gn}(t)} \ket{gn} + \sqrt{\mu_n(t)} \mathrm{e}^{\mathrm{i}\left(\theta_n + \varphi_{fn}(t)\right)} \ket{fn} \right)
\end{equation}
$\mu_n(t)$ is the occupancy of $\ket{fn}$ relative to the occupancy of Fock state $n$, $\varphi_{gn}(t)$ the phase evolution of the $\ket{gn}$ state and $\varphi_{fn}(t)$ the phase evolution of the $\ket{fn}$ state less the target phase shift $\theta_n$. 

To derive the connection between the no-jump trajectory and the path-dependence errors for $f\rightarrow e$ decay, we rewrite the Lindblad Master equation from \cref{equ:Lindblad_Master_equ} as:
\begin{equation}
	\label{equ:Lindblad_ET_violations_decay}
	\frac{\mathrm{d}\hat{\rho}}{\mathrm{d}t} = -\mathrm{i}\left[\hat{H}_\mathrm{eff}, \hat{\rho}\right] + \Gamma_{f\rightarrow e} \hat{L}_{f\rightarrow e} \hat{\rho} \hat{L}_{f\rightarrow e}^\dag
\end{equation}
with the effective (non-hermitian) Hamiltonian $\hat{H}_\mathrm{eff} = \hat{H}_\mathrm{rot} -\mathrm{i}/2 \Gamma_{f\rightarrow e} \hat{L}_{f\rightarrow e}^\dag \hat{L}_{f\rightarrow e}$ and $\hat{H}_\mathrm{rot}$ from \cref{equ:H_rot_gf_SNAP}. Apart from the $f\rightarrow e$ decay rate, all other noise rates are set to zero. The solution of the Schr\"odinger equation corresponding to $\hat{H}_\mathrm{eff}$ is labeled as $\ket{\psi_\mathrm{eff}(t)}$ and is identical to the time evolution in \cref{equ:no-jump-trajectory} up to a first order correction in $\Gamma_{f\rightarrow e} t$:
\begin{equation}
    \ket{\psi_\mathrm{eff}(t)} = \ket{\psi(t)} + \mathcal{O}(\Gamma_{f\rightarrow e} t)
\end{equation}

We assume that the $\chi$-matching condition is fulfilled, so the Lindblad operator regarding $f\rightarrow e$ decay from \cref{equ:lindblad_operator_f_e_decay} simplifies to $\hat{L}_{f\rightarrow e} = \ketbra{e}{f}$. By introducing $\hat{\rho}^\prime=\ketbra{e}{e}\hat{\rho}\ketbra{e}{e}$ and $\hat{\rho}_0=\hat{\rho}-\hat{\rho}^\prime$, we can rewrite \cref{equ:Lindblad_ET_violations_decay} as:
\begin{subequations}
	\begin{align}
		\label{equ:ET_violations_decay_rho0}
		\frac{\mathrm{d}\hat{\rho}_0}{\mathrm{d}t} &= -\mathrm{i}\left[\hat{H}_\mathrm{eff}, \hat{\rho}_0\right] \\
		\label{equ:ET_violations_decay_rho_prime}
		\frac{\mathrm{d}\hat{\rho}^\prime}{\mathrm{d}t} &= \Gamma_{f\rightarrow e} \hat{L}_{f\rightarrow e} \hat{\rho}_0 \hat{L}_{f\rightarrow e}^\dag
	\end{align}
\end{subequations} 
The solution for \cref{equ:ET_violations_decay_rho0} is given by $\hat{\rho}_0(t) = \ketbra{\psi_\mathrm{eff}(t)}{\psi_\mathrm{eff}(t)}$. With \cref{equ:ET_violations_decay_rho_prime} and $\hat{\rho}^\prime (0)=0$, we get:
\begin{subequations}
	\begin{align}
		\hat{\rho}^\prime (T) &= \Gamma_{f\rightarrow e} \int_0^T  \ket{e}\braket{f}{\psi_\mathrm{eff}(t)} \braket{\psi_\mathrm{eff}(t)}{f}\bra{e}  \,dt =\\
		\label{equ:final_rho_prime_approx}
		&= \Gamma_{f\rightarrow e} \int_0^T  \ket{e} \braket{f}{\psi (t)} \braket{\psi (t)}{f}\bra{e}  \,dt + \mathcal{O}\left((\Gamma_{f\rightarrow e} T)^2\right)
	\end{align}
\end{subequations}

The probability to be in the first excited state, but not in the target state is the mean squared overlap error for path dependence regarding $f\rightarrow e$ decay:
\begin{equation}
	\label{equ:ET_violations_decay_error}
	 \underset{\substack{\vec{c} \in \mathbb{C}^L \text{ s.t.}\\ \norm{c}=1}}{\mathrm{avg}} \Delta \mathcal{F}^{(2)}_\mathrm{PD,f\rightarrow e} =  \underset{\substack{\vec{c} \in \mathbb{C}^L \text{ s.t.}\\ \norm{c}=1}}{\mathrm{avg}} P(e) -  \underset{\substack{\vec{c} \in \mathbb{C}^L \text{ s.t.}\\ \norm{c}=1}}{\mathrm{avg}} P(|\psi_{\mathrm{tgt},e}\rangle )
\end{equation}

The average $e$ state occupancy is given by:
\begin{equation}
	 \underset{\substack{\vec{c} \in \mathbb{C}^L \text{ s.t.}\\ \norm{c}=1}}{\mathrm{avg}} P(e) = \underset{\substack{\vec{c} \in \mathbb{C}^L \text{ s.t.}\\ \norm{c}=1}}{\mathrm{avg}} \sum_n \bra{en} \hat{\rho}^\prime (T) \ket{en} 
	= \frac{1}{L}\Gamma_{f\rightarrow e} \sum_n \int_0^T \mu_n(t) \,dt + \mathcal{O}\left((\Gamma_{f\rightarrow e} T)^2\right)
\end{equation}

The averaged probability to be in the target state $\ket{\psi_{\mathrm{target},e}}$ (see \cref{equ:target_state_e}) turns out to equal:
\begin{subequations}
	\begin{align}
		&\underset{\substack{\vec{c} \in \mathbb{C}^L \text{ s.t.}\\ \norm{c}=1}}{\mathrm{avg}}  P(\ket{\psi_{\mathrm{target},e}} ) =\underset{\substack{\vec{c} \in \mathbb{C}^L \text{ s.t.}\\ \norm{c}=1}}{\mathrm{avg}}  \bra{\psi_{\mathrm{target},e}} \hat{\rho}^\prime (T) \ket{\psi_{\mathrm{target},e}}\\
		&\qquad =  
		\frac{1}{L(L+1)}\Gamma_{f\rightarrow e}\sum_{n,n^\prime}(1+\delta_{n,n^\prime})\int_0^T \sqrt{\mu_n(t)\mu_{n^\prime}(t)}\cos (\varphi_{fn^\prime}(t)-\varphi_{fn}(t) ) + \mathcal{O}\left((\Gamma_{f\rightarrow e} T)^2\right)
	\end{align}
\end{subequations}

\subsubsection{Path-independence violations for transmon dephasing for the optimized $ge$ and $gf$ SNAP protocol}
\label{supp:ET_violations_dephasing}
We will now repeat the analysis performed in \cref{supp:path_independence_violations_decay} for transmon dephasing. We perform the calculations here for the $gf$ SNAP protocol, but the result is also valid for the $ge$ protocol. All noise contributions, apart from transmon dephasing, are set to zero. The path-independence violations for the unoptimized SNAP protocols are negligible compared to the coherent error, and are not further considered.

As the following calculations are conceptually simple, but the intermediate steps are quite long expression, only the relevant ideas and the main results are presented.

From the evolution of the quantum state defined in \cref{equ:no-jump-trajectory}, we get the coherent time evolution operator $U(t)$:
\begin{equation}
	U(t) = \sum_n
	\begin{pmatrix}
		\sqrt{1-\mu_n(t)} \mathrm{e}^{\mathrm{i}\varphi_{gn}(t)} & 0 & - \sqrt{\mu_n(t)} \mathrm{e}^{-\mathrm{i}\left(\theta_n + \varphi_{fn}(t)\right)} \\
		0 & 0 & 0\\
		\sqrt{\mu_n(t)} \mathrm{e}^{\mathrm{i}\left(\theta_n + \varphi_{fn}(t)\right)} & 0 & \sqrt{1-\mu_n(t)} \mathrm{e}^{-\mathrm{i}\varphi_{gn}(t)}
	\end{pmatrix}
	\otimes \ketbra{n}{n}
\end{equation}
The Lindblad Master equation is given by:
\begin{equation}
	\frac{\mathrm{d}\hat{\rho}}{\mathrm{d}t} = -\mathrm{i}\left[\hat{H}_\mathrm{eff}, \hat{\rho}\right] + \Gamma_{ff} \hat{L}_{ff} \hat{\rho} \hat{L}_{ff}^\dag
\end{equation}
with the effective (non-hermitian) Hamiltonian $\hat{H}_\mathrm{eff} = \hat{H}_\mathrm{rot} -\mathrm{i}/2 \Gamma_{ff}\hat{L}_{ff}^\dag \hat{L}_{ff}$. Without loss of generality, we split $\hat{\rho}$ into $\hat{\rho}_0$ and $\rho^\prime$ with $\hat{\rho}^\prime(0)=0$ and:
\begin{subequations}
	\begin{align}
		\label{equ:LBM_rho_0_ET_dephasing}
		\frac{\mathrm{d}\hat{\rho}_0}{\mathrm{d}t} &= -\mathrm{i}\left[\hat{H}_\mathrm{eff}, \hat{\rho}_0\right] \\
		\label{equ:LBM_rho_prime_ET_dephasing}
		\frac{\mathrm{d}\hat{\rho}^\prime}{\mathrm{d}t} &= -\mathrm{i}\left[\hat{H}_\mathrm{eff}, \hat{\rho}^\prime\right] + \Gamma_{ff} \hat{L}_{ff} \hat{\rho} \hat{L}_{ff}^\dag
	\end{align}
\end{subequations}

\cref{equ:LBM_rho_0_ET_dephasing} is identical with \cref{equ:ET_violations_decay_rho0}. 

From the initial condition for $\hat{\rho}^\prime$ and from \cref{equ:LBM_rho_prime_ET_dephasing} follows that $\hat{\rho}^\prime$ is of order $\Gamma_{ff}T$ and we can rewrite \cref{equ:LBM_rho_prime_ET_dephasing} as:
\begin{equation}
	\frac{\mathrm{d}\hat{\rho}^\prime}{\mathrm{d}t} = -\mathrm{i}\left[\hat{H}_\mathrm{rot}, \hat{\rho}^\prime\right] + \Gamma_{ff} \hat{L}_{ff} \ketbra{\psi(t)}{\psi(t)} \hat{L}_{ff}^\dag + \mathcal{O}\left((\Gamma_{ff} T)^2\right)
\end{equation}
To solve this equation, we transform it into the frame comoving with $\hat{H}_\mathrm{rot}$:
\begin{equation}
	\label{equ:rho_prime_ET_dephasing}
	\frac{\mathrm{d}\hat{\widetilde{\rho}}^\prime}{\mathrm{d}t} =   \Gamma_{ff} \hat{U}^\dag(t)\hat{L}_{ff} \ketbra{\psi(t)}{\psi(t)} \hat{L}_{ff}^\dag \hat{U}(t) + \mathcal{O}\left((\Gamma_{ff} T)^2\right)
\end{equation}
with $\hat{\widetilde{\rho}}^\prime (t) = \hat{U}^\dag(t) \hat{\rho}^\prime (t) \hat{U}(t)$. Integrating \cref{equ:rho_prime_ET_dephasing} in time and transforming the result into the original frame gives us the final density matrix. We assume that the end points of the \enquote{no-jump} trajectory $\ket{\psi(t)}$ are optimized with our scheme, so $\mu(T)$ is 1 and $\varphi_{fn}(T)$ is 0. 

We can finally calculate the matrix elements:
\begin{subequations}
	\begin{align}
		\begin{split}
			&\underset{\substack{\vec{c} \in \mathbb{C}^L \text{ s.t.}\\ \norm{c}=1}}{\mathrm{avg}}  \langle \psi_{\mathrm{tgt},g}|\hat{\rho}^\prime (T) | \psi_{\mathrm{tgt},g} \rangle =\underset{\substack{\vec{c} \in \mathbb{C}^L \text{ s.t.}\\ \norm{c}=1}}{\mathrm{avg}}   \frac{\Gamma_{ff}}{L(L+1)} \sum_{n,n^\prime} (1+\delta_{n,n^\prime})  \int_0^T \sqrt{\mu_n(t)\mu_{n^\prime}(t)} \times \\
			& \qquad\times \sqrt{1-\mu_n (t)}\sqrt{1-\mu_{n^\prime}(t)} \cos(\varphi_{fn}(t)-\varphi_{fn^\prime}(t)+\varphi_{gn}(t)-\varphi_{gn^\prime}(t))  \,\mathrm{d}t 
		\end{split} \\
		&\underset{\substack{\vec{c} \in \mathbb{C}^L \text{ s.t.}\\ \norm{c}=1}}{\mathrm{avg}}  \langle \psi_{\mathrm{tgt},g}|\hat{\rho}_0 (T) | \psi_{\mathrm{tgt},g} \rangle \approx 0 \\
		& \underset{\substack{\vec{c} \in \mathbb{C}^L \text{ s.t.}\\ \norm{c}=1}}{\mathrm{avg}}  \langle \psi_{\mathrm{tgt},f}|\hat{\rho}^\prime (T) | \psi_{\mathrm{tgt},f} \rangle = \frac{\Gamma_{ff}}{L(L+1)} \sum_{n,n^\prime} (1+\delta_{n,n^\prime}) \int_0^T \mu_n(t) \mu_{n^\prime}(t) \,\mathrm{d}t \\
		&\underset{\substack{\vec{c} \in \mathbb{C}^L \text{ s.t.}\\ \norm{c}=1}}{\mathrm{avg}}  \langle \psi_{\mathrm{tgt},f}|\hat{\rho}_0 (T) | \psi_{\mathrm{tgt},f} \rangle \approx 1 - \frac{\Gamma_{ff}}{L} \int_0^T \sum_n \mu_n (t) \,\mathrm{d}t
	\end{align}
\end{subequations}

The mean squared overlap error of the path-independence violations regarding transmon dephasing is then given by:
\begin{subequations}
	\label{equ:ET_violations_dephasing_error}
	\begin{align}
		\Delta \mathcal{F}^{(2)}_\mathrm{PD,ff} &\approx \underset{\substack{\vec{c} \in \mathbb{C}^L \text{ s.t.}\\ \norm{c}=1}}{\mathrm{avg}} \left(\underbrace{P(g) + P(f)}_{=1} - P(| \psi_{\mathrm{tgt},g} \rangle) - P(| \psi_{\mathrm{tgt},f} \rangle)\right) \\
		\begin{split}
		&\approx 1 - \underset{\substack{\vec{c} \in \mathbb{C}^L \text{ s.t.}\\ \norm{c}=1}}{\mathrm{avg}}  \langle \psi_{\mathrm{tgt},g}|\hat{\rho}^\prime (T) | \psi_{\mathrm{tgt},g} \rangle -\underset{\substack{\vec{c} \in \mathbb{C}^L \text{ s.t.}\\ \norm{c}=1}}{\mathrm{avg}}  \langle \psi_{\mathrm{tgt},f}|\hat{\rho}_0 (T) | \psi_{\mathrm{tgt},f} \rangle - \\
		& \qquad - \underset{\substack{\vec{c} \in \mathbb{C}^L \text{ s.t.}\\ \norm{c}=1}}{\mathrm{avg}}  \langle \psi_{\mathrm{tgt},f}|\hat{\rho}^\prime (T) | \psi_{\mathrm{tgt},f} \rangle 
		\end{split}
	\end{align}
\end{subequations}

\subsection{Summary of all error contributions}
\cref{tab:summary_comparison_SNAP_protocols} summarizes the different SNAP gate protocols discussed in the main text in Fig.~(3) and their best performances.

\begin{table}[]
    \centering
    \begin{tabular}{c|c|c}
        SNAP protocol &  optimal gate time $\chi T_\mathrm{opt}$& optimal mean squared overlap error \\
        \hline
        unoptimized ge protocol & 6.5$\pi$ & \num{0.0951}\\
        optimized ge protocol & 2.7$\pi$ (optimization limit) & \num{0.0262} \\
        optimized ge protocol with error detection & 2.7$\pi$ (optimization limit) & \num{0.0172} \\
        unoptimized gf protocol with error detection  & 12.0$\pi$& \num{0.0227} \\
        optimized gf protocol with error detection & 4.0$\pi$ & \num{0.0099}
    \end{tabular}
    \caption{Summary of the different SNAP protocols and their optimal working points shown in the main text in Fig.~(3)}
    \label{tab:summary_comparison_SNAP_protocols}
\end{table}

\section{Analysis of the interference measurements}
\label{supp:interference_measurements}
In this section, we summarize how the phase errors of the SNAP gate are linked to the interference populations.
We write the final state of the SNAP gate as:
\begin{equation}
    \ket{\psi_\mathrm{final}} = \sum_n c_{n,\mathrm{final}}\mathrm{e}^{\mathrm{i}(\theta_n + \Delta \theta_n)} \ket{gn} + \ket{e} \otimes \ket{\psi_e}
\end{equation}
with $c_{n,\mathrm{final}}$ as the amplitude vector of the final cavity state, $\theta_n$ as the target phase shift of the SNAP operation, and $\Delta \theta_n$ as the phase errors. $\ket{\psi_e}$ is the part of the final state, where the transmon is in its excited state $\ket{e}$. In our experiments, the initial cavity state only includes real amplitudes $c_n$ (we only used real $\alpha$ for the interference measurements and, therefore, the final amplitude vector is considered to be real throughout this section. The elements of the final amplitude vector are connected to the population measurements by:
\begin{equation}
    c_{n,\mathrm{final}} = \sqrt{P(g,n)}
\end{equation}
Performing the interference measurement yields the quantity:
\begin{equation}
    P_\epsilon (g,n) = \abs{\bra{gn}\hat{D}(\epsilon) \ket{\psi_\mathrm{final}}}^2
\end{equation}
The displacement operator $\hat{D}(\epsilon)$ can be written as \cite{cahill1969ordered}:
\begin{equation}
    D(\epsilon) = \sum_{mn} \underbrace{\sqrt{\frac{n!}{m!}} \epsilon ^{m-n} \mathrm{e}^{-\abs{\epsilon}^2/2} L_n^{(m-n)} (\abs{\epsilon}^2)}_{\eqqcolon d_{mn}(\epsilon)}\ketbra{m}{n}
\end{equation}
with $L_n^{(k)}(x)$ as the generalized Laguerre polynomials. 
Note, that the we only used real displacements $\epsilon$ in our experiments. Therefore, we expand $P_\epsilon (g,n)$ as:
\begin{equation}
    \label{equ:P_eps_msmt}
    P_\epsilon (g,n) = \sum_{n1} \abs{d_{nn_1}}^2  P(g, n_1) + 2 \sum_{\substack{n_1,n_2\\n_1<n_2}} d_{nn_1}(\epsilon) d_{nn_2} (\epsilon) \sqrt{P(g,n_1)P(g,n_2)} \cos(\theta_{n_2} + \Delta \theta_{n_2} - \theta_{n_1} - \Delta \theta_{n_1} )
\end{equation}
With the measured populations $P(g,n)$ and the interference populations $P_\epsilon (g,n)$, we receive a nonlinear system of equations with the phase errors $\Delta \theta_n$ as the unknowns. Solving this nonlinear equation system results in the phase errors. However, \cref{equ:P_eps_msmt} is not very sensitive with respect to the phase errors and measurement errors limit the performance of this analysis. The resulting phase errors show huge fluctuations and the precision of this analysis is not better than \num{0.2}--\SI{0.3}{\radian}. 

We estimate an upper bound for the phase errors which go beyond the theoretically expected phase errors in the following way: We assume that the deviations between the experimentally measured and simulated interference populations (see Fig.~2(i, j) in the main text) can be fully attributed to additional phase errors. Using error propagation, the size of these phase errors turns out to be \SI{0.24}{\radian}, which agrees well to the size of the fluctuations seen in our analysis. Unfortunately, this precision is not sufficient to demonstrate experimentally the improvement of the phase errors resulting from our optimization protocol.

\section{System parameters}
\label{sec:system_parameters}

\begin{table}
	\begin{center}
		\begin{tabular}{l|r}
			parameter name & numerical value \\
			\hline
			qubit $ge$ frequency $\omega_{ge}$ & $2\pi\times\SI{4.092820}{\giga \hertz}$  \\
			cavity frequency $\omega_c$&  $2\pi\times\SI{4.484628}{\giga \hertz}$  \\
			dispersive shift $\chi$ & $2\pi\times\SI{486.1(3)}{\kilo\hertz}$ \\
			Kerr constant $K$ & $2\pi\times\SI{699(6)}{\hertz}$ \\
			correction to the dispersive shift $\chi^\prime$ & $2\pi\times\SI{0.97(7)}{\kilo \hertz}$  \\
			qubit $ge$ $T1$ & \SI{110(1)}{\micro \second} \\
			qubit $ge$ $T2$ Ramsey& \SI{48(2)}{\micro \second} \\
			qubit $ge$ $T2$ Hahn echo & \SI{105(4)}{\micro \second} \\
			cavity $T1$ & \SI{1.00(2)}{\milli \second}
		\end{tabular}
	\end{center}
	\caption{System parameters.}
	\label{tab:SNAP-system-parameters}
\end{table}

\cref{tab:SNAP-system-parameters} summarizes the system parameters of our setup. To smooth the switching on and off of the SNAP pulses, the pulse functions in Eqs.~(3) and (5) are multiplied with an envelope function $\mathrm{env}(t)$. The envelope function is normalized, such that $\int_0^T \mathrm{env}(t)=T$. In our experiments, we used the envelope function:
\begin{equation}
    \mathrm{env}(t) = \frac{\beta}{\beta T - \pi} \times
    \begin{cases}
		\frac{1}{2}\left(1-\cos(\beta t)\right) &\text{if } 0\leq t \leq \pi/\beta \\
		1 &\text{if } \pi/\beta \leq t \leq T-\pi/\beta \\
		\frac{1}{2}\left(1+\cos(\beta t)\right) &\text{if } T-\pi/\beta \leq t\leq T \\
		0 & \text{else}
	\end{cases}
\end{equation}
where $\beta$ is a smoothening coefficient set to $2\pi/(0.2 T)$ for all of our experiments.

The second stage of our SNAP gate is implemented with a fast and unselective $\pi$ pulse with Gaussian envelope and a duration of \SI{48}{\nano \second}.

\section{Robustness of the SNAP gates with respect to parameter errors}

\begin{figure}
    \centering
    \includegraphics[width=0.8\textwidth]{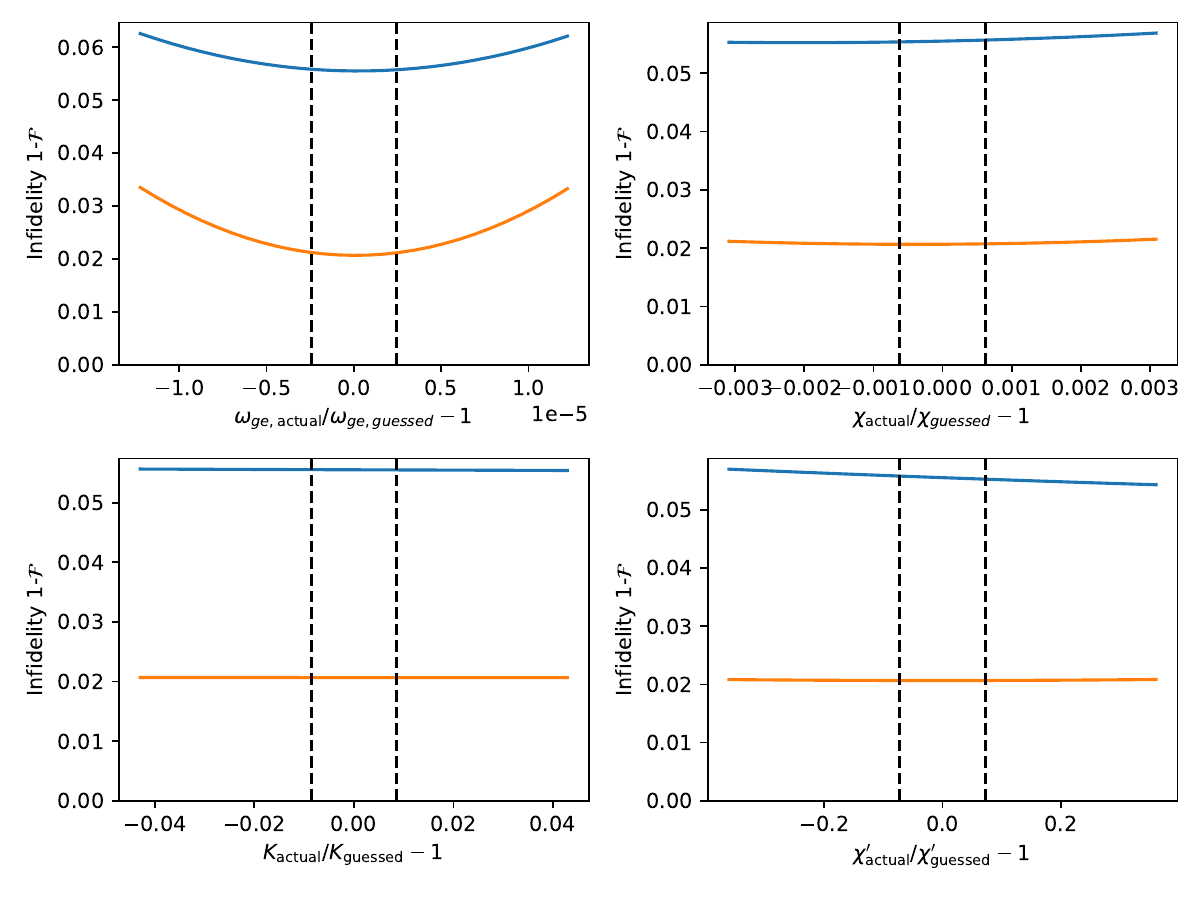}
    \caption{Theoretically predicted infidelity of the (un)optimised SNAP gates with respect to errors in the system parameters. In every subfigure the actual system parameter is varied with respect to the set of parameters the SNAP pulses were computed for. The typical error range of the parameters is indicated by vertical dashed lines. Considered target operation: $(0,\pi,0,\pi,0,\pi)$ with $\chi_\mathrm{guessed} T=3\pi$.}
    
    \label{fig:parameter_errors}
\end{figure}

The SNAP gates' performance is robust with respect to errors in the system parameters within typical error ranges. In \cref{fig:parameter_errors} we show how the performance of the SNAP gates depends on errors in the system parameters. The parameters not shown include the loss rates for which the performance loss is even smaller.

\section{Comparison to GRAPE optimised pulses}
\begin{figure}
    \centering
    \includegraphics[width=0.3\textwidth]{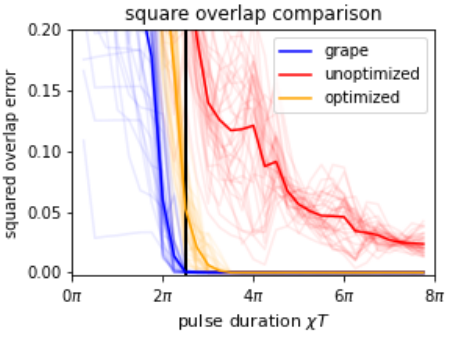}
    \caption{Coherent error of the unoptimised SNAP pulses (red), our optimised SNAP protocol (orange), and GRAPE optimised pulses (blue). The target operation $(0,\theta_1,\theta_2,0,0,0)$ is applied to the first six Fock states. Solid lines show averages of the 36 different gates optimised while the transparent lines show the results of individual optimisations. The threshold at which GRAPE reaches a fidelity error of \SI{10e-4} is marked by the black vertical line.}
    \label{fig:GRAPE-comparison}
\end{figure}


In \cref{fig:GRAPE-comparison} we compare the ability of GRAPE and our optimisation scheme to minimize coherent errors. Similar to our optimisation scheme, GRAPE is able to almost completely suppress coherent errors. GRAPE is able to reach slightly lower gate times ($T\approx 2.5 \pi /\chi$) compared to our optimisation scheme ($T\approx 3\pi/\chi$). This is sort of expected given the GRAPE pulse has a much higher number of free parameters. However, our optimisation scheme uses only 36 optimisation parameters in this case  and reaches almost the same fidelity compared to GRAPE, which is based on 5000 or more parameters. In addition our optimisation approach runs much faster and is conveniently parametrized in terms of distinct frequency components. Furthermore, we foresee that our scheme can be implemented by a feedback loop interacting directly with the actual experimental setup, see Appendix M for more details.

\section{Wigner tomography and fidelity estimation}

The cavity parity is mapped onto the transmon state by a Ramsey sequence with a wait-time of $\pi/\chi$. The actual used wait time in the experiment is a bit lower due to the finite duration of the transmon $\pi/2$ pulses and is calibrated by applying the Wigner function measurement to different Fock states 0 to 9 and observing where the parity readout is maximised. The experimentally shorter wait time agrees with a simulation including the finite-duration of the Ramsey pulses. The $T_1$ of the transmon leads to a Wigner function not centered perfectly at zero. In order to remove such a non-physical bias half of the statistics is measured by changing the phase of the second Ramsey pulse by $\pi$, interchanging the roles of $\ket{g}$ and $\ket{e}$ in the readout. The readouts are then combined to the parity value $P=P_{g1}-P_{e2}$ where subscripts 1 and 2 denote the first and second half of the data taking.

We define the fidelity as the overlap between the measured quantum state $\hat{\rho}_\mathrm{out}$ and the target state $\ket{\psi_\mathrm{target}}$. Written in terms of Wigner functions, the fidelity equals:
\begin{equation}
    \mathcal{F} = \bra{\psi_\mathrm{target}} \hat{\rho}_\mathrm{out} \ket{\psi_\mathrm{target}} = \frac{4}{\pi}\int \mathrm{d}^2 \alpha \, W^\mathrm{out}(\alpha)W^\mathrm{target}(\alpha)
\end{equation}
with $\mathrm{d}^2\alpha$ as a surface element in phase space and $W^\mathrm{out}(\alpha)$ and $W^\mathrm{target}(\alpha)$ as the Wigner functions of the output and target state. Both Wigner functions are normalised such that $\int \mathrm{d}^2\alpha W(\alpha) = \pi/2$.

To estimate the fidelity for our pulses, we perform a Wigner tomography of the final cavity state as described above and compute the Wigner function of the target state. For a discretized phase space the fidelity equals:
\begin{equation}
    \mathcal{F} = \frac{4}{\pi} \Delta A \sum_{ij} W_{ij}^\mathrm{out}(\alpha_{ij})W_{ij}^\mathrm{target}(\alpha_{ij})
\end{equation}
with $\Delta A$ as the size of the surface element. Assuming that every pixel of the measured Wigner function has a constant error $\Delta W$, the fidelity error turns out to equal:
\begin{equation}
    \Delta \mathcal{F} = \sqrt{\frac{4}{\pi}\Delta A} \Delta W
\end{equation}
In our experiment we estimate $\Delta W$ from the standard deviation of the fluctuations in the outer rim of multiple Wigner tomographies, where the function is expected to be zero. Estimating $\Delta W$ as $0.055$ the fidelity error equals $\Delta \mathcal{F}=0.013$.

\section{Pulse optimisation in the experiment}
In this work we optimise the SNAP pulses in simulations. However, we foresee that our approach can be realised by a feedback loop directly interacting with the actual experimental setup. A key element of our approach is that every coherent error can be controlled and corrected individually by their respective pulse parameter. By measuring the operators $\hat{\sigma}_{\alpha_n}\otimes \ketbra{n}{n}$ and $\hat{\sigma}_{\alpha_n+\pi/2}\otimes \ketbra{n}{n}$ one can measure the transversal and longitudinal errors individually for each Fock mode (see \cref{equ:terminal_state_coordinates}). $\hat{\sigma}_{\alpha}$ is the $\hat{\sigma}_x$ Pauli operator rotated by the angle $\alpha$ around the z axis: $\hat{\sigma}_{\alpha}=\hat{\sigma}_x \cos \alpha  + \hat{\sigma}_y \sin \alpha$. Also the phase errors can be measured individually by performing the interference measurements discussed in FIG.~2(a) in the main text. Therefore, all coherent errors and the respective correction parameters can be determined from experimentally accessible quantities. This allows us to optimise the pulses directly in the experiment, which would e.g.~help to address calibration errors or pulse distortions.

\section{Simulation time}
Our optimization scheme is very fast and requires much less numerical effort than GRAPE. For target operations with four modes our scheme takes about a minute or less to converge when starting from the unoptimized pulses while running on the CPU of our laptop. If one has optimized pulses for a certain parameter set and these parameters have changed slightly, one can reapply the optimization starting from the already pre-trained pulses and we expect much faster convergence speeds. 

\end{appendix}

\bibliography{main.bib}